\documentclass[acmsmall,nonacm]{acmart}

\usepackage[figuresright]{rotating}
\usepackage{stfloats} 
\usepackage{enumitem}
\usepackage{multirow}
\usepackage{listings}
\usepackage{url}
\usepackage{amsfonts}
\usepackage{mathtools}
\usepackage{amsthm}
\usepackage{color}
\usepackage{algorithm}  
\usepackage{makecell}
\usepackage{subcaption}
\usepackage{fontawesome5}
\usepackage{colortbl}
\usepackage{framed}
\usepackage{lscape}

\definecolor{lightgreen}{rgb}{0.25, 0.63, 0.4375}
\definecolor{darkblue}{rgb}{0.02, 0.16, 0.49}
\definecolor{codegray}{gray}{.9}
\definecolor{darkgreen}{rgb}{0, 0.5, 0}
\definecolor{darkred}{rgb}{0.72,0.04,0.04}

\definecolor{jsonkey}{RGB}{0,0,180}    
\definecolor{jsonstring}{RGB}{0,0,0}    
\definecolor{jsoncomment}{RGB}{0,128,0} 
\definecolor{paramvalue}{RGB}{184,92,0} 

\lstdefinelanguage{myjson}{
    basicstyle=\ttfamily\small,
    showstringspaces=false,
    breaklines=true,
    morestring=[b]",
    morecomment=[l]{//},
    keywordstyle=\color{jsonkey},
    stringstyle=\color{jsonstring},
    commentstyle=\color{jsoncomment}\ttfamily,
    morekeywords={"body"}
}

\lstdefinestyle{gocomment}{
        language=go,
        numbers=left,    
        numbersep=4pt,
        keywordstyle=\color{black},
        moredelim=[is][\color{darkgreen}]{!!}{!!}
}

\lstdefinelanguage{json}{
  basicstyle=\ttfamily\small,
  numbers=none,
  numberstyle=\tiny,
  breaklines=true,
  showstringspaces=false,
  stringstyle=\color{blue},
  morestring=[b]",
  literate=
   *{:}{{{\color{black}:}}}{1}
    {,}{{{\color{black},}}}{1}
    {[}{{{\color{black}[}}}{1}
    {]}{{{\color{black}]}}}{1}
    {\{}{{{\color{black}\{}}}{1}
    {\}}{{{\color{black}\}}}}{1}
}

\definecolor{answergray}{gray}{0.90}
\definecolor{promptgray}{gray}{0.95}
\definecolor{prompttitle}{gray}{0.30}

\newenvironment{answerboxenv}[1][answergray]
{
  
  \vspace{4pt}%
  \MakeFramed{\advance\hsize-\width \FrameRestore}%
  \noindent %
  \ignorespaces %
}
{
  \unskip %
  \endMakeFramed
  \vspace{4pt}
}

\newcommand{\answerbox}[2][answergray]{%
  \begin{answerboxenv}[#1]%
  \ignorespaces #2%
  \end{answerboxenv}
}

\newenvironment{promptbox}[1]
{
  
  \vspace{10pt}
  \MakeFramed{\advance\hsize-\width \FrameRestore}
  \noindent
    \leftskip=0pt
    \rightskip=0pt
  \raggedright   

  \vspace*{-4pt}%
  {
    \noindent\hspace*{-6pt}%
    \setlength{\fboxsep}{4pt}
    \colorbox{prompttitle}{%
        \parbox{\dimexpr\linewidth+8pt-2\fboxsep\relax}{%
          \color{white}\bfseries #1
        }%
    }
    \par\vspace{4pt}%
  }
}
{
  \endMakeFramed
  \vspace{4pt}
}

\newcommand{\prompt}[2]{%
  \begin{promptbox}{#1}
  #2
  \end{promptbox}
}

\definecolor{commentcolor}{RGB}{31,135,29}

\lstdefinestyle{custom}{
  commentstyle=\color{commentcolor},
  moredelim=[il][\textcolor{commentcolor}]{\#}
}

\newcommand{\framework}{\textsc{APIDiffer}}

\acmDOI{10.1145/3798219}
\acmYear{2026}
\acmJournal{PACMPL}
\acmVolume{10}
\acmNumber{OOPSLA1}
\acmArticle{111}
\acmMonth{4}
\received{2025-10-10}
\received[accepted]{2026-02-17}

\begin{document}

\title{When Specifications Meet Reality: Uncovering API Inconsistencies in Ethereum Infrastructure}

\author{Jie Ma}
\orcid{0009-0007-6311-520X}
\affiliation{%
  \institution{Beihang University}
  \city{Beijing}
  \country{China}
}
\affiliation{%
  \institution{Zhongguancun Laboratory}
  \city{Beijing}
  \country{China}
}
\email{majie2023@buaa.edu.cn}

\author{Ningyu He}
\orcid{0000-0002-9980-7298}
\affiliation{%
  \institution{The Hong Kong Polytechnic University}
  \city{Hong Kong SAR}
  \country{China}
}
\email{ningyu.he@polyu.edu.hk}

\author{Jinwen Xi}
\orcid{0000-0002-7504-3457}
\affiliation{%
  \institution{Zhongguancun Laboratory}
  \city{Beijing}
  \country{China}
}
\email{xijw@zgclab.edu.cn}

\author{Mingzhe Xing}
\orcid{0000-0002-2065-9852}
\affiliation{%
  \institution{Zhongguancun Laboratory}
  \city{Beijing}
  \country{China}
}
\email{xingmz@zgclab.edu.cn}

\author{Liangxin Liu}
\orcid{0000-0002-1331-7643}
\affiliation{%
  \institution{Beihang University}
  \city{Beijing}
  \country{China}
}
\email{by2439116@buaa.edu.cn}

\author{Jiushenzi Luo}
\orcid{0009-0003-1415-1133}
\affiliation{%
  \institution{Beijing Institute of Technology}
  \city{Beijing}
  \country{China}
}
\affiliation{%
  \institution{Zhongguancun Laboratory}
  \city{Beijing}
  \country{China}
}
\email{3120235248@bit.edu.cn}

\author{Xiaopeng Fu}
\orcid{0009-0008-2907-080X}
\affiliation{%
  \institution{Beijing Institute of Technology}
  \city{Beijing}
  \country{China}
}
\affiliation{%
  \institution{Zhongguancun Laboratory}
  \city{Beijing}
  \country{China}
}
\email{fuxp2024@zgclab.edu.cn}

\author{Chiachih Wu}
\orcid{0009-0009-6006-3306}
\affiliation{%
  \institution{Amber Group}
  \city{Hong Kong SAR}
  \country{China}
}
\email{chiachih.wu@ambergroup.io}

\author{Haoyu Wang}
\orcid{0000-0003-1100-8633}
\affiliation{%
  \institution{Huazhong University of Science and Technology}
  \city{Wuhan}
  \country{China}
}
\email{haoyuwang@hust.edu.cn}

\author{Ying Gao}
\orcid{0000-0001-8992-651X}
\affiliation{%
  \institution{Beihang University}
  \city{Beijing}
  \country{China}
}
\affiliation{%
  \institution{Zhongguancun Laboratory}
  \city{Beijing}
  \country{China}
}
\email{gaoying@buaa.edu.cn}
\authornote{Ying Gao and Yinliang Yue are the corresponding authors.}

\author{Yinliang Yue}
\orcid{0000-0002-8417-2234}
\affiliation{%
  \institution{Zhongguancun Laboratory}
  \city{Beijing}
  \country{China}
}
\email{yueyl@zgclab.edu.cn}
\authornotemark[1]

\begin{abstract}
The Ethereum ecosystem, which secures over \$381 billion in assets, fundamentally relies on client APIs as the sole interface between users and the blockchain. 
However, these critical APIs suffer from widespread implementation inconsistencies, which can lead to financial discrepancies, degraded user experiences, and threats to network reliability.
Despite this criticality, existing testing approaches remain manual and incomplete: they require extensive domain expertise, struggle to keep pace with Ethereum's rapid evolution, and fail to distinguish genuine bugs from acceptable implementation variations.
We present {\framework}, the first specification-guided differential testing framework designed to automatically detect API inconsistencies across Ethereum's diverse client ecosystem.
{\framework} transforms API specifications into comprehensive test suites through two key innovations: 
(1) specification-guided test input generation that creates both syntactically valid and invalid requests enriched with real-time blockchain data, and 
(2) specification-aware false positive filtering that leverages large language models to distinguish 
genuine bugs from acceptable variations.
Our evaluation across all 11 major Ethereum clients reveals the pervasiveness of API bugs in production systems.
{\framework} uncovered 72 bugs, with 90.28\% already confirmed or fixed by developers, including one critical error in the official specifications themselves.
Beyond these raw numbers, {\framework} achieves up to 89.67\% higher code coverage than existing tools and reduces false positive rates by 37.38\%. The Ethereum community's response validates our impact: developers have integrated our test cases, expressed interest in adopting our methodology, and escalated one bug to the official Ethereum Project Management meeting.
By making {\framework} open-source, we enable continuous validation of Ethereum client API implementations, thereby strengthening the foundational integrity of the entire Ethereum ecosystem.
\end{abstract}

\ccsdesc[500]{Software and its engineering~Software testing and debugging}
\ccsdesc[300]{Security and privacy~Software security engineering}

\keywords{Ethereum, Ethereum Client API, Differential Testing, Large Language Models}

\renewcommand{\shortauthors}{Jie Ma, Ningyu He et al.}

\maketitle

\section{Introduction}
\label{sec:intro}
As one of the most active blockchain platforms, 
Ethereum~\cite{buterin2013ethereum} hosts a diverse ecosystem of applications, 
including decentralized finance~\cite{defi}, stable coins~\cite{stablecoin}, NFTs~\cite{nft}, and Web3 games~\cite{game}.
At the time of writing, the total market capitalization of Ethereum has reached up to \$381.49 billion~\cite{CoinMarket}.
Currently, the Ethereum network comprises approximately 12K mainnet nodes~\cite{nodes_account}. 
Following \textit{The Merge}~\cite{history}, each Ethereum node must integrate both an execution layer (EL) client and a consensus layer (CL) client to operate. 
Deploying such nodes requires substantial resources (CPU, disk, and network bandwidth)~\cite{run_node}, making Ethereum client API services the primary way for most users to interact with Ethereum, rather than running their own nodes.
For instance, when Ethereum users submit a transaction, it must be signed with their private key, a process typically handled by a cryptocurrency wallet (\textit{e.g.,} MetaMask~\cite{metamask}) or Web3 libraries (\textit{e.g.,} Web3.py~\cite{Web3py}, ethers.js~\cite{ethers.js}).
Despite the convenience these tools provide, all of them ultimately rely on the Ethereum client API at the lower level to interact with the Ethereum network~\cite{rpc_example}. 

\setlength{\fboxsep}{0pt}
\setlength{\fboxrule}{0.5pt}

\begin{figure}[t]
    \centering
    \begin{subfigure}[b]{0.75\textwidth}
    \fcolorbox{gray!10}{white}{\includegraphics[width=1\textwidth]{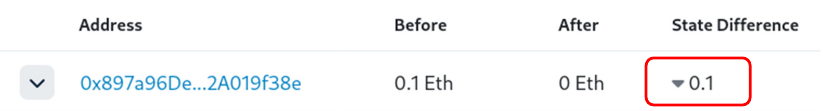}}
    \caption{Query the transfer value of a specific transaction on Etherscan; the value shown in the \textit{State Difference} section is 0.1 ETH.}
    \label{fig:example:0.1}
    \end{subfigure}
    \hfill 
    \begin{subfigure}[b]{0.75\textwidth}
    \fcolorbox{gray!10}{white}{\includegraphics[width=1\textwidth]{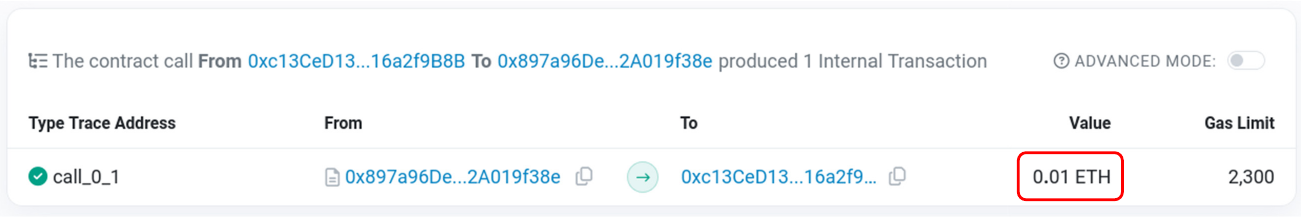}}
    \caption{Query the same transaction on Etherscan; the value shown in the \textit{Internal Transactions} section and the real transfer value on the chain is 0.01 ETH.}
    \label{fig:example:0.01}
    \end{subfigure}
    \vspace{-0.1in}
    \caption{A motivating example of a real-world Ethereum client API bug on Etherscan. In this case, the transfer value displayed in the browser (0.1 ETH in Fig.\ref{fig:example:0.1}) is incorrectly shown as ten times the actual amount recorded on the blockchain (0.01 ETH in Fig.\ref{fig:example:0.01}). The root cause is that Etherscan uses a faulty Ethereum client API provided by Erigon~\cite{erigon_bug}.}
    \vspace{-0.1in}
    \label{fig:mot_example}
    \Description{title}
\end{figure}

Given this central role in the Ethereum ecosystem, ensuring the correctness of client API implementations becomes paramount.
Bugs in these implementations can have severe consequences for the entire ecosystem, including, but not limited to, inconsistencies when requesting blockchain data (\textit{e.g.,} wrong account balances) and poor user experience (\textit{e.g.,} different response from identical request). Such issues threaten the overall consistency and reliability of the Ethereum network. Fig.~\ref{fig:mot_example} illustrates a real-world Ethereum client API bug observed on Etherscan, the most widely used Ethereum browser~\cite{etherscan}.  
Etherscan is a website that allows users to view and search blockchain data, such as transactions and blocks~\cite{etherscan}.
In this case, the actual transfer value of a transaction (0.01 ETH) was mistakenly displayed in the browser as 0.1 ETH, because Etherscan relies on a faulty Ethereum client API implementation~\cite{erigon_bug}. 
This seemingly small discrepancy could mislead users, disrupt financial transactions, and undermine trust in Ethereum ecosystem,
highlighting the critical need for comprehensive testing of API implementations. 

This necessity raises a critical question: \textit{How can we systematically test the Ethereum client API implementations?} 
The decentralized nature of Ethereum, with multiple independent client implementations, presents both a challenge and an opportunity. 
While traditional API testing approaches~\cite{arcuri_tool_2024,api_survey,evomaster,restest} target single providers, Ethereum's inherent diversity~\cite{diversity}, multiple implementations of the same protocol, makes differential testing~\cite{mckeeman1998differential} particularly suitable. By comparing responses across different implementations, we can identify inconsistencies without requiring predefined oracles.

\begin{table}[t]
\caption{Comparison of {\framework} with existing Ethereum client API testing tools in terms of supported targets and capabilities.}
\label{table:challenges}
\begin{tabular}{c|c|c|c}
\toprule
\multirow{2}{*}{\textbf{Tool}} &
  \multirow{2}{*}{\textbf{\begin{tabular}[c]{@{}c@{}}Supported\\ Testing Target\end{tabular}}} &
  \multirow{2}{*}{\textbf{\begin{tabular}[c]{@{}c@{}}Challenge \#1\\ Test Input Generation\end{tabular}}} &
  \multirow{2}{*}{\textbf{\begin{tabular}[c]{@{}c@{}}Challenge \#2 \\ Bug Identification\end{tabular}}} \\
            &                     &               &                      \\ \hline
EtherDiffer & EL client API       & DSL           & human                \\
rpctestgen  & EL client API       & template      & human                \\
\rowcolor{codegray} 
APIDiffer   & EL \& CL client API & specification & specification \& LLM \\
\bottomrule 
\end{tabular}
\vspace{-0.1in}

\end{table}

After conducting a comprehensive literature review and tool search, we found that, there are only two tools specifically designed for Ethereum client API testing, \textit{i.e.,} EtherDiffer~\cite{kim2023etherdiffer} and \texttt{rpctestgen}~\cite{rpctg}.
\textit{Applying differential testing to Ethereum client API implementations faces several challenges}, which can be summarized in two aspects in Table~\ref{table:challenges}. 
On the one hand, \textit{it is challenging to generate test inputs for various Ethereum client API implementations.}
EtherDiffer adopts a DSL-based approach to generate test inputs for EL client APIs, while \texttt{rpctestgen} is the official testing tool maintained by Ethereum developers, relying on manually crafted test cases.
These two approaches rely on domain-specific languages~\cite{kim2023etherdiffer} or manually crafted test cases~\cite{rpctg}, both of which require significant expertise and human effort. Moreover, these methods struggle to keep pace with the rapid evolution of Ethereum and suffer from scalability issues, making them inadequate for testing.
On the other hand, \textit{it is challenging to identify real bugs from testing results.} 
Although differential testing does not require predefined oracles, identifying bugs from observed inconsistencies remains difficult due to the complex structure and variability of the responses.
Besides, inevitable environmental differences, permitted response inconsistencies, and semantically equivalent yet syntactically different outputs often lead to false positives. These false positives require extra human effort for bug identification and future fix.

\textbf{Our work.} 
To bridge this critical gap in Ethereum's infrastructure testing, we present {\framework}, a specification-guided differential testing framework. Unlike existing approaches that rely on manual effort or outdated templates, {\framework} automatically synthesizes comprehensive test requests directly from official API specifications, covering both EL and CL clients for the first time.
{\framework} introduces two key innovations. First, it generates not only syntactically valid requests but also semantically meaningful ones by extracting real blockchain data to ensure tests interact with actual on-chain entities rather than non-existent addresses or blocks. Second, it dramatically reduces false positives through intelligent filtering that combines specification-aware heuristics with large language models to distinguish genuine bugs from acceptable implementation variations.
Our evaluation reveals {\framework}'s impact: across all 11 major Ethereum clients that collectively secure \$381 billion in assets, {\framework} uncovered 72 bugs, 90.28\% of which have been confirmed or fixed by developers, even including one critical error in the official API specifications themselves. {\framework} achieves up to 89.67\% higher code coverage than state-of-the-art tools while reducing false discovery rates by 37.38\%.

\textbf{Contribution.} In summary, we make the following contributions:
\begin{itemize}[leftmargin=*]
    \item We present {\framework}, the first comprehensive testing framework for both EL and CL Ethereum client APIs, introducing specification-guided test generation that reduces manual effort.
    \item We develop an automated bug identification approach that reduces false positives by 37.38\% through semantic equivalence analysis.
    \item We demonstrate {\framework}'s effectiveness by uncovering 72 real-world bugs (90.28\% confirmed or fixed) across all major Ethereum clients, achieving up to 89.67\% higher coverage than baselines.
    \item We contribute to the Ethereum ecosystem with open-source tool and actionable insights, with one bug escalated to official project management discussions.
\end{itemize}

\section{Background}

\begin{figure}[t]
    \centering
    \includegraphics[width=0.8\textwidth]{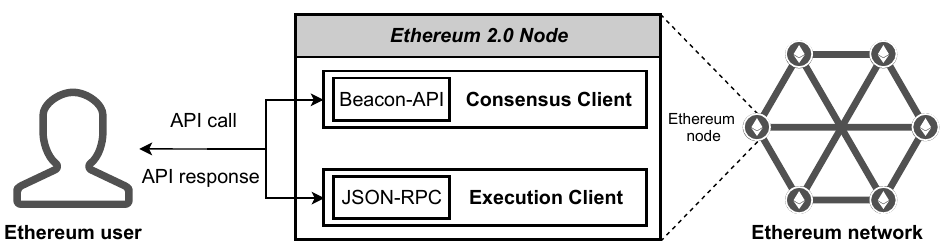}
    \vspace{-0.1in}
    \caption{The relationship among users, Ethereum nodes, Ethereum blockchain network, and EL/CL clients along with their exposed APIs.}
    \label{fig:node}
    \Description{title}
\end{figure}

\subsection{Ethereum Node}
\label{section:background:node}
As illustrated in Fig.~\ref{fig:node}, \textit{Ethereum node} is serving as the gate between external users and the Ethereum blockchain network. Following Ethereum's transition from the proof-of-work to the proof-of-stake consensus mechanism in 2022, known as \textit{The Merge}~\cite{merge}, each Ethereum node now consists of two components, \textit{i.e.,} the \textit{execution layer (EL)} client and the \textit{consensus layer (CL)} client~\cite{node}.
\begin{itemize}[leftmargin=*]
    \item \textbf{EL client.} EL client, formerly known as \textit{the Eth1 client}, is responsible for execution-related tasks, \textit{e.g.,} transaction validation, propagation, and processing, as well as state management and smart contract execution within the Ethereum Virtual Machine (EVM)~\cite{evm}.
    \item \textbf{CL client.} CL client is also known as \textit{the Beacon Node} or \textit{the Eth2 client}, which implements the proof-of-stake consensus mechanism that enables the Ethereum network to reach agreement. 
    According to the specification~\cite{consensus-specs}, once 32 ETH is staked, the CL client can function as a validator, participating in block validation and proposal. This incentive mechanism rewards nodes for contributing to the consensus process while imposing financial penalties, such as slashing, for malicious or non-compliant behavior.
\end{itemize}

\begin{table}[t]
\caption{Details of existing Ethereum mainnet EL clients and CL clients.}
\vspace{-0.1in}
\label{table:target:clients}
\resizebox{0.65\linewidth}{!}{%
\begin{tabular}{cccccc}
\toprule
\textbf{Type} & \textbf{Project} & \textbf{Share} & \textbf{Language} & \textbf{Stars} & \textbf{Version}\\
\midrule
& Geth~\cite{go-ethereum}             & 41\%                & Golang            & 49.3k                 & v1.15.12                                           \\
& Nethermind~\cite{nethermind}       & 38\%                & C\#               & 1.4k                  & v1.31.11                                           \\
& Besu~\cite{besu}           & 16\%                & Java              & 1.7k                  & 25.6.0                                             \\
& Erigon~\cite{erigon}          & 3\%                 & Golang            & 3.4k                  & v3.0.11                                            \\
\multirow{-5}{*}{EL client} & Reth~\cite{reth}             & 2\%                 & Rust              & 4.8k                  & v1.4.8                                             \\
\midrule
& Lighthouse~\cite{lighthouse}       & 42.71\%             & Rust              & 3.2k                  & v7.0.1                                             \\
& Prysm~\cite{prysm}           & 30.91\%             & Golang            & 3.6k                  & v6.0.4                                             \\
& Teku~\cite{teku}          & 13.86\%             & Java              & 708                   & 25.5.0                                             \\
                                   & Nimbus~\cite{nimbus}          & 8.74\%              & Nim               & 594                   & v25.6.0                                            \\
                                   & Lodestar~\cite{lodestar}         & 2.67\%              & Typescript        & 1.3k                  & v1.31.0 \\
\multirow{-6}{*}{CL client} & Grandine~\cite{grandine}         & 1.04\%              & Rust              & 221                   & 1.1.1   \\         
\bottomrule                                
\end{tabular}
}
\end{table}
\label{section:background:diversity}

Ethereum is known for its emphasis on decentralization and security, where a key factor contributing to this is the client diversity~\cite{diversity}, which plays a critical role in enhancing fault tolerance and reducing the risk of consensus failures~\cite{re_diversity}. 
Specifically, client diversity refers to \textit{the presence of multiple independent implementations of the Ethereum protocol, both execution and consensus, that must be functionally equivalent but developed independently}. 
As shown in Table~\ref{table:target:clients}, the Ethereum mainnet currently are dominated by five EL clients, \textit{i.e.,} Geth~\cite{go-ethereum}, Nethermind~\cite{nethermind}, Besu~\cite{besu}, Erigon~\cite{erigon}, Reth~\cite{reth}, and six CL clients, \textit{i.e.,} Lighthouse~\cite{lighthouse}, Prysm~\cite{prysm}, Teku~\cite{teku}, Nimbus~\cite{nimbus}, Lodestar~\cite{lodestar}, Grandine~\cite{grandine}, implemented in various programming languages.
Despite their diverse design patterns and implementations, \textit{these clients are supposed to achieve same functionalities and behave consistently}.

\subsection{Ethereum Client API}
\label{section:background:nodeapi}
The only interface exposed to users by Ethereum nodes is the Ethereum client API.
As maintaining an Ethereum node is not trivial for normal users due to hardware and software requirements~\cite{run_node}, most Ethereum users interact with the blockchain through Ethereum client APIs exposed by third-party Ethereum nodes.
Specifically, there are two types of client APIs: EL client APIs (\textit{e.g.,} JSON-RPC API~\cite{jsonrpc_api}) and CL client APIs (\textit{e.g.,} Beacon API~\cite{beacon_api}).
JSON-RPC APIs define a set of methods for interacting with the EL client, enabling operations such as transaction submission, state queries, and block retrieval. Beacon APIs allow users to query and participate in the Ethereum Beacon Chain. They are standardized via shared specifications~\cite{jsonrpc_api,beacon_api}, widely supported across multiple clients and actively maintained. Thanks to this standardization, users and blockchain applications, like crypto wallets, can consistently achieve various functionalities without needing to focus on the underlying API implementations or operate their own Ethereum node.
While there are other exposed APIs, for example, GraphQL exposed by some EL clients~\cite{graphql} and various client-specific APIs~\cite{customapi}, they are inconsistently implemented and lack broad support. Accordingly, this work focuses on these two well-supported sets of APIs (\textit{i.e.,} JSON-RPC API and Beacon API), ensuring our analysis remains relevant and comparable across different client implementations.

\subsection{Differential Testing}
Differential testing is a widely-adopted technique to identify potential bugs by comparing the behaviors or outputs of multiple independent implementations of the same functionality without requiring oracles~\cite{mckeeman1998differential}. 
When performing differential testing, an identical input is provided to each implementation, and their outputs or behaviors are compared. Any discrepancies may indicate unexpected bugs.
Differential testing first requires generating test requests for the target programs to execute. Then, during execution, the runtime information will be collected for comparison and further bug reporting. Differential testing is particularly effective for uncovering semantic or logical bugs that do not exhibit explicit erroneous behaviors, such as crashes or assertion failures.
\section{Challenges \& Solutions}
In this section, we first present the challenges of performing differential testing targeting Ethereum client APIs through  a concrete motivating example in \S\ref{subsec:challenge}.
Then, we provide a high-level overview of our solutions in \S\ref{subsec:solution}.

\subsection{Motivating Example \& Challenges}
\label{subsec:challenge}

Fig.~\ref{fig:mot_example} illustrated in \S\ref{sec:intro} is a real-world Ethereum client API bug in Etherscan~\cite{geth_example}, the most well-known Ethereum explorer, which can be regarded as a service that accesses on-chain data through Ethereum client APIs and visualizes the corresponding data.
This bug occurs when querying a transaction\footnote{Transaction Hash: 0x96dfd56413baf7ee53483d5d6788ea5faa621ee6bb7d4b8808408d1f07f72bf8} on Etherscan, where the transfer value is displayed differently on the external (0.1 ETH in Fig.~\ref{fig:example:0.1}) and internal (0.01 ETH in Fig.~\ref{fig:example:0.01}) user views.
The root cause of this bug lies in Etherscan's reliance on two APIs provided by Erigon~\cite{erigon}, an EL client, separately in its inner and outer user interfaces. 
In the outer layer, Etherscan uses \texttt{trace\_transaction} to retrieve transaction metadata. Because \texttt{trace\_transaction} lacks support for capturing the beacon root contract call, the retrieved transfer value is incorrect. In the inner user interface, Etherscan utilizes \texttt{debug\_traceTransaction} to parse the transaction metadata and obtain the correct transfer value.

\begin{sloppypar}
A similar issue has also been observed in Reth~\cite{reth}, another EL client~\cite{reth_bug}, where \texttt{debug\_traceTransaction} fails to execute the beacon root contract call, while \texttt{trace\_transaction} behaves correctly. 
We underline that \textit{severe impacts could be brought by such subtle differences}.
For example, users may mistakenly believe that the transaction has transferred 0.1 ETH, leading to misinformed decisions in trust, investment, or trading. Generally speaking, this bug also undermines user confidence in the integrity and robustness of the whole blockchain ecosystem, opens up possibilities for malicious exploitation such as fraud or public opinion manipulation, and reveals underlying consistency vulnerabilities in Ethereum client API implementations. 
\end{sloppypar}
Considering its potential huge impact and the diversity of clients, we plan to conduct a comprehensive and thorough testing of various Ethereum client API implementations using differential testing. We have summarized two key challenges.

\noindent\textbf{Challenge \#1: Generating test inputs for various Ethereum client API.} 
Testing first requires generating test inputs. Under the context of this paper, a test input is a request to the target API, primarily consisting of the API name and its associated parameters. Currently, certain methods have been developed for API test input generation~\cite{kim2023etherdiffer,rpctg,enhance_rest,Adaptive_REST,evomaster,erigon_rpc_test}, which can be broadly classified into three categories:
1) \textit{Manual composing.} This category often indicates developers directly writing test cases for APIs, typically through unit tests. 
While this approach provides a high level of control and precision in testing specific API functions~\cite{erigon_rpc_test}, it is inherently time-consuming, error-prone, and lacks scalability for complex systems like Ethereum.
2) \textit{Automatic composing.} Fully automatic methods, particularly those leveraging machine learning models or Large Language Models (LLMs),
aim to intelligently explore the vast search space of API operations and parameters~\cite{enhance_rest,Adaptive_REST,evomaster}. 
While these methods significantly reduce manual effort, they often struggle with precision; specifically, the automatically generated requests frequently fail to satisfy the rigorous syntactic or semantic constraints required by complex protocols.
3) \textit{Semi-automatic composing.} Methods in this category typically utilize domain-specific languages (DSLs) or templates to help developers generate API test cases~\cite{kim2023etherdiffer,rpctg}. These approaches automate certain steps of test case generation while leaving other parts to developers. It usually means finding a balance between the first two approaches. However, they still impose a significant burden on developers to maintain DSLs or templates as the underlying protocol evolves.

\noindent\textbf{Challenge \#2: Identifying true positive bugs from inconsistencies.} 
Although differential testing does not require predefined oracles, identifying actual bugs from observed inconsistencies remains difficult due to the complex structure and variability of the responses. Specifically, the following scenarios may lead to unnecessary false positives.
\begin{itemize}[leftmargin=*]
    \item \textbf{Differences in testing environments and node states.} Variations in runtime conditions and synchronization status across nodes can introduce inconsistencies that are not indicative of actual bugs. For instance, the response of certain APIs (\textit{e.g.,} \texttt{getSyncingStatus}) depends on the internal state of the Ethereum node, and nodes across the network may not be synchronized at the same block height. If a node is still syncing, it may be unable to serve specific API requests. In such cases, different responses from syncing nodes are expected, and treating these as bugs would lead to false positives.
    \item \textbf{Allowed inconsistencies in response.} Certain fields in the response are allowed to differ across nodes and should not be considered indicators of bugs. For example, fields like \texttt{peer\_id}, which represent a node’s public key, are inherently unique and can not be treated as inconsistencies.
    \item \textbf{Semantically equivalent but syntactically different responses.} Since clients are implemented in different programming languages, they may return responses that differ in format or structure but convey the same semantic meaning. For example, the \texttt{message} field in error responses is a string that provides details when an API call fails, where different clients may produce different error strings that are semantically equivalent. When calling \texttt{publishBlindedBlockV2} with invalid input, CL client Nimbus returns the message ``Unable to decode data'', while another CL client Prysm responds with ``could not decode request body into...''. Despite the different wording, both messages convey the same underlying error.
\end{itemize}

\noindent\textbf{Limitations of current tools.} 
To the best of our knowledge, only a few tools are available for testing Ethereum client API implementations, \textit{i.e.,} EtherDiffer~\cite{kim2023etherdiffer} and \texttt{rpctestgen}~\cite{rpctg}. 
While client teams may maintain their own internal test suites for their API implementations, these test suites are typically not designed to cover all clients. The most obvious limitation is that different test suites may be written in different languages, migrating one to another raises scalability challenges, thus beyond our scope. 
Next, we argue that all existing tools have limitations in addressing our proposed challenges.

Specifically, EtherDiffer is a differential testing framework for EL client APIs which introducing a DSL to generate API requests.
However, CL client APIs follows a different design paradigm, \textit{i.e.,} they are RESTful APIs rather than JSON-RPC (see \S\ref{section:background:nodeapi}). Moreover, the DSL-based approach heavily relies on human expertise and lacks scalability, especially given the rapid and ongoing evolution of the Ethereum protocol specification. Designing a new DSL for CL client APIs poses significant challenges, as we mentioned in \textbf{Challenge \#1}.
Furthermore, EtherDiffer lacks a systematic mechanism for identifying false positives in testing results, where all observed deviations are treated as bugs and require tedious and error-prone manual intervention and investigation, which hinders solving \textbf{Challenge \#2}.
As for \texttt{rpctestgen}, it is part of the official Ethereum API testing framework maintained by Ethereum developers~\cite{hive}. The test inputs generated by \texttt{rpctestgen} are hard-coded in the \texttt{testgen} package, making it subject to the limitations described in \textbf{Challenge \#1}. Regarding \textbf{Challenge \#2}, similarly, any deviation between the observed responses and expected responses is treated as a test failure~\cite{hive_panel}, resulting in numerous false positives that require manual confirmation.

\subsection{Solutions}
\label{subsec:solution}
To address these fundamental challenges in Ethereum API testing, {\framework} introduces two key innovations that transform the testing landscape from manual, expertise-dependent processes to automated, specification-driven methodologies.

To tackle \textbf{Challenge \#1}, we propose a \textit{specification-guided request generation method.} 
Rather than relying on outdated domain knowledge or incomplete manual test suites, {\framework} leverages Ethereum's well-maintained API specifications as the primary source for systematic test input generation. Our approach operates at two levels: first, we employ \textit{syntax-oriented request generation} to automatically produce both syntactically valid and invalid requests directly from JSON schema definitions, eliminating the need for expert knowledge about complex API structures and constraints. Second, we introduce \textit{fact-based semantics-aware request generation} that extracts real blockchain data through auxiliary API calls to populate test parameters with meaningful values, ensuring that generated requests can successfully interact with live client implementations rather than failing due to non-existent blockchain entities.

Regarding \textbf{Challenge \#2}, we propose a \textit{specification-aware false positive filtering method} to distinguish genuine implementation bugs from acceptable client variations. {\framework} develops a multi-layered filtering approach that understands the nuanced nature of blockchain API responses. We first apply heuristic-based filtering to eliminate known environmental differences and implementation-specific behaviors. More importantly, we leverage large language models to automatically analyze API specifications and classify response fields into three behavioral categories: \textit{must-identical} (fields that should be identical across all clients), \textit{may-divergent} (fields that may legitimately differ), and \textit{must-divergent} (fields that are expected to differ). This specification-aware classification enables {\framework} to focus on true inconsistencies while filtering out false positives, particularly for semantically equivalent responses that differ only in syntactic representation.

\begin{figure*}[t] 
    \centering 
    \includegraphics[width=1\linewidth]{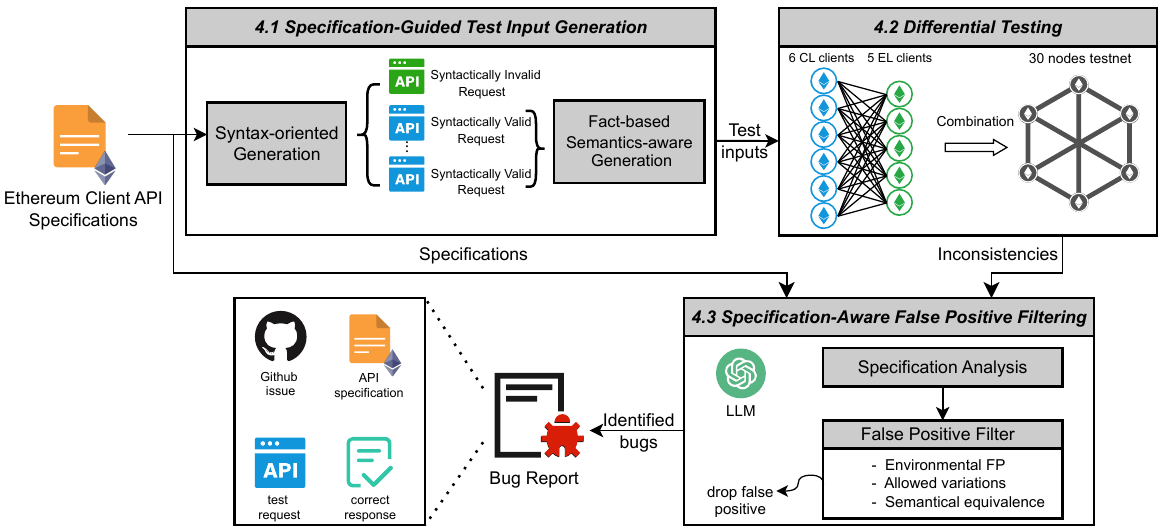} 
    \vspace{-0.1in}
    \caption{The workflow overview of \framework. The inputs include EL and CL client API specifications, and the output is the detailed bug report.}
    \vspace{-0.2in}
    \label{fig:overview} 
    \Description{The overview of \framework. }
\end{figure*}

\section{Approach}
This section begins with an overview of {\framework} in \S\ref{subsec:overview}. Subsequently, we present a detailed examination of its three core phases from \S\ref{subsec:testcase_generate} to \S\ref{subsec:bug}, respectively.

\subsection{Overview}
\label{subsec:overview}
{\framework} introduces a novel specification-guided differential testing framework that systematically detects API bugs across Ethereum's diverse client ecosystem. 
As illustrated in Fig.~\ref{fig:overview}, {\framework} operates through three interconnected phases that transform Ethereum's well-documented specifications into comprehensive testing capabilities. 
First, {\framework} automatically generates syntactically valid and invalid test inputs from JSON-RPC~\cite{exec_api_github} and Beacon-API~\cite{beacon_api} specifications, then enriches them with semantically meaningful parameters through fact-based semantics-aware request generation approach that extracts live blockchain data. 
Second, {\framework} deploys these inputs across a controlled local testnet comprising all major client combinations, executing same requests simultaneously to capture response inconsistencies. 
Finally, {\framework} employs specification-aware false positive filtering that automatically classifies response fields based on expected behavior patterns and applies LLM-based semantic equivalence detection to distinguish genuine implementation bugs from acceptable variations, generating actionable bug reports. 
These three phases are detailed in the following.

\subsection{Phase I: Specification-Guided Test Input Generation}
\label{subsec:testcase_generate}

To address Challenge \#1, {\framework} leverages the well-documented specifications that accompany both EL and CL client APIs~\cite{exec_api,beacon_api}. These specifications define essential metadata including request and response structures, parameter constraints, and functionality descriptions. Based on this foundation, {\framework} implements a fully automated specification-guided test input generation method consisting of two complementary stages.
First, \textit{syntax-oriented request generation} creates both syntactically valid and invalid requests directly from specification schemas to test API correctness and robustness. 
However, specifications impose only syntactic constraints without semantic validation, for example, the Line 7 of Fig.~\ref{fig:spec:api} requires the \texttt{Address} field to be a hexadecimal string, a randomly generated value may not correspond to an actual blockchain address. 
Second, \textit{fact-based semantics-aware request generation} addresses this limitation by enriching those syntactically valid requests with semantically meaningful values retrieved from live blockchain data, thereby facilitating more effective testing.

\subsubsection{Syntax-oriented Request Generation.}
\label{subsubsec:json:generator}
{\framework} employs a schema-based approach~\cite{hypothesis-jsonschema} to automatically create both syntactically valid and invalid API requests. Specifically, for each parameter of APIs included in the specification, the schema defines its type, validation pattern, and acceptable values. 
For example, the \texttt{eth\_getBalance} API in Fig.~\ref{fig:spec:api} specifies two parameters with distinct schema, where \texttt{Address} requires a hexadecimal string matching a specific pattern (Line 6 -- Line 8), while \texttt{Block} accepts three alternative formats: number, tag, or hash (Line 10 -- Line 19).

\begin{sloppypar}
\noindent \textbf{Valid Request Generation.} 
To generate syntactically correct requests satisfying the constraints, instead of reinventing a wheel from scratch, we adopted the established approach from \texttt{hypothesis-jsonschema}~\cite{hypothesis-jsonschema}, which extends the \texttt{Hypothesis} property-based testing framework~\cite{hypothesis} to automatically derive generation strategies from schema definitions.
Specifically, by taking advantage of the \texttt{from\_schema} interface, {\framework} could convert each schema of parameters into the corresponding data generator. Concrete test inputs are then synthesized via the \texttt{example} function~\cite{hypothesis-example}. 
Fig.~\ref{fig:eth_getBalance_valid} demonstrates a generated valid request, where both the \texttt{method} and formatted \texttt{params} conform to the corresponding schema defined in the API specification.
\end{sloppypar}

\noindent \textbf{Invalid Request Generation.} To test the implementation robustness, {\framework} generates requests that \textit{violate} constraints while preserving the overall API structure (fields like \texttt{id}, \texttt{method}, and \texttt{jsonrpc} in Fig.~\ref{fig:eth_getBalance_invalid}). 
We introduce three categories of violations: (1) requests containing undefined fields, (2) requests lacking required fields, and (3) requests containing values that break type or pattern constraints. 
Specifically, to introduce (1) undefined fields, we leverage the \texttt{additionalProperties} configuration in \texttt{from\_schema} to deliberately inject fields beyond those defined in the specification.
To simulate (2) missing fields, we randomly remove those \textit{required} parameters in generated test cases. 
Finally, to (3) violate constraints, we randomly replace parameter values with data of another type. 
As illustrated in Fig.~\ref{fig:eth_getBalance_invalid}, this approach produces a request the first field is an undefined field.

\lstdefinelanguage{customjson}{
    basicstyle=\ttfamily\small\color{black},
    breaklines=true,
    showstringspaces=false,
    numbers = none,
    frame=single,
    morecomment=[l]{//},
    commentstyle=\color{jsoncomment}\ttfamily,
    keywords={name,summary,params,type,pattern,anyOf,enum,result,id,jsonrpc,method,required,schema,title,description},
    keywordstyle=\color{blue},
    stringstyle=\color{black},
    literate={"x-consistency-policy"}{{\textcolor{blue}{"x-consistency-policy"}}}1
}

\begin{figure}[t]
    \centering
    \begin{lstlisting}[
    language=customjson,
    breaklines=true,
    numbers=left,
    columns=fullflexible,
    xleftmargin=1.5em,
    xrightmargin=1em,
    ]
{ 
  "name": "eth_getBalance",
  "summary": "Returns the balance of the account of given address.",
  "params": [ {
      "name": "Address", "required": true,
      "schema": {
        "title": "hex encoded address", "type": "string",
        "pattern": "^0x[0-9a-fA-F]{40}$" } }, {
      "name": "Block", "required": true,
      "schema": {
        "title": "Block number, tag, or block hash",
        "anyOf": [ {
            "title": "Block number", "type": "string",
            "pattern": "^0x(0|[1-9a-f][0-9a-f]*)$" }, {
            "title": "Block tag", "type": "string",
            "enum": ["earliest", "finalized", "safe", "latest", "pending"],
            "description": "..." }, {
            "title": "Block hash", "type": "string",
            "pattern": "^0x[0-9a-f]{64}$" } ] } } ],
  "result": {
    "name": "Balance",
    "schema": {
      "title": "hex encoded unsigned integer", "type": "string",
      "pattern": "^0x(0|[1-9a-f][0-9a-f]*)$"
// This field is added by the LLM to filter out false positives    
      "x-consistency-policy", "must-identical"} } 
}
    \end{lstlisting}
    \vspace{-0.1in}
    \caption{The specification of Ethereum client API \texttt{eth\_getBalance}.} 
    \vspace{-0.3in}
    \label{fig:spec:api} 
    \Description{The specification of API \texttt{eth\_getBalance}.}
\end{figure}

\begin{figure}[t]
    \centering 
    \begin{subfigure}[t]{\linewidth}
        \begin{lstlisting}[
            language=customjson, 
            frame=single,   
            breaklines=true, 
            breakatwhitespace=false,
            columns=fullflexible,
            escapeinside={(*@}{@*)},
            xleftmargin=1em,      
            xrightmargin=1em,
        ]
{
  "id": 1,
  "jsonrpc": "2.0",
  "method": "eth_getBalance",
  "params": [ 
    (*@\color{paramvalue} "\string\ud982\string\udd65": [ [], 10000000.0, true, "0xfe97" ]@*),
    (*@\color{paramvalue} "0xffffeeee...ffffeeeeffff" @*)]
}
    \end{lstlisting}
        \caption{Example of a syntactically invalid test request of API \texttt{eth\_getBalance}, where undefined field is included.}
        \vspace{0.1in}
        \label{fig:eth_getBalance_invalid}
    \end{subfigure}
    \vspace{0.1in}
    \begin{subfigure}{0.46\linewidth}
        \begin{lstlisting}[
        language=customjson,
        columns=fullflexible,
        escapeinside={(*@}{@*)},
        xleftmargin=1em,      
        ]
{
  "id": 1,
  "jsonrpc": "2.0",
  "method": "eth_getBalance",
  "params": [
    (*@\color{paramvalue}\ttfamily\small   "0xffffeeee...ffffeeeeffff"@*),
    "latest"
  ]
}
        \end{lstlisting}
    \caption{Example of a syntactically valid but semantically invalid request of API \texttt{eth\_getBalance}, where the fields \texttt{Address} and \texttt{Block} conform to the specification, but the \texttt{Address} does not exist.}
    \label{fig:eth_getBalance_valid}
    \end{subfigure}
    \hfill
    \begin{subfigure}{0.46\linewidth}
        \begin{lstlisting}[
            language=customjson,
            columns=fullflexible,
            escapeinside={(*@}{@*)},
            xrightmargin=1em,
        ]
{
  "id": 1,
  "jsonrpc": "2.0",
  "method": "eth_getBalance",
  "params": [
    "0x9411d2eb...4EBb1c8ADA7B",
    "latest"
  ]
}
        \end{lstlisting}
    \caption{Example of a semantically valid test input of API \texttt{eth\_getBalance} generated based on Fig.~\ref{fig:eth_getBalance_valid}, where the value of the required field \texttt{Address} is fetched using \texttt{eth\_getTransactionReceipt}.}

    \label{fig:eth_getBalance_sem_valid}
    \end{subfigure}

    \vspace{-0.1in}
    \caption{Test requests generated by \framework, with orange highlights indicating syntactically or semantically invalid fields.} 
    \Description{title.}
    \vspace{-0.28in}
    \label{fig:api:example} 
\end{figure}

\begin{table}[t]
\caption{Parameter types and corresponding source APIs for fact-based semantics-aware request generation.}
\label{table:seman:mut}
\resizebox{0.9\columnwidth}{!}{%
\begin{tabular}{cccc}
\toprule
\textbf{Parameter}   & \textbf{Parameter Options}   & \textbf{Layer} & \textbf{Available APIs}                                                                                                                                                    \\
\midrule
\multirow{3}{*}{Block}                & block number                 & EL             & \begin{tabular}[c]{@{}c@{}}eth\_getBlockByNumber\\ eth\_getBlockReceipts\end{tabular}                                                                                       \\ 

\cline{2-4}
                     & block hash                   & EL             & \begin{tabular}[c]{@{}c@{}}eth\_getBlockByHash\\ eth\_getBlockReceipts\end{tabular}                                                                                         \\
\midrule
Address              & address identifiers          & EL             & \begin{tabular}[c]{@{}c@{}}eth\_getTransactionByHash\\ eth\_getTransactionReceipt\end{tabular}                                                                              \\
\midrule
\multirow{8}{*}{Transaction}         & full\_transaction            & EL             & \begin{tabular}[c]{@{}c@{}}eth\_blockNumber\\ eth\_getBlockByNumber\\ eth\_getBlockByHash\\ eth\_getTransactionByBlockHashAndIndex\\ eth\_getTransactionByHash\end{tabular} \\
\cline{2-4}
                     & transaction\_hash            & EL             & \begin{tabular}[c]{@{}c@{}}eth\_sendTransaction\\ eth\_getBlockReceipts\\ eth\_getFilterChanges\end{tabular}                                                                \\
\cline{2-4}
                     & transaction\_receipt         & EL             & eth\_getTransactionReceipt                                                                                                                                                  \\
\cline{2-4}
                     & transaction\_count            & EL             & \begin{tabular}[c]{@{}c@{}}eth\_getBlockTransactionCountByHash\\ eth\_getBlockTransactionCountByNumber\end{tabular}           
                     \\
\cline{2-4}                     
                     & nonce                        & EL             & eth\_getTransactionCount                                                                                                                                                    \\
\midrule
Filter               & filter\_id                   & EL             & eth\_newFilter                                                                                                                                                              \\
\midrule
\multirow{2}{*}{state\_id}           
                     & slot                         & CL             & getBlockHeader                                                                                                                                                              \\
                     & stateroot                    & CL             & getBlockHeader
                     \\
\midrule                     
epoch                & epoch identifier             & CL             & getBlockHeader                                                                                                                                                              \\
\midrule
finalized\_stateroot & finalized state identifiers & CL             & getBlockHeader                                                                                                                                                              \\
\midrule
\multirow{2}{*}{block\_id}            & slot                         & CL             & getBlockHeader                                                                                                                                                              \\
                     & stateroot                    & CL             & getBlockHeader                                                                                                                                                              \\
\midrule
\multirow{2}{*}{validator\_id}        & validator\_index             & CL             & postStateValidatorIdentities                                                                                                                                                \\
                     & validator\_pubkey            & CL             & postStateValidatorIdentities                                                                                                                                                \\
\midrule
committee\_index     & committee identifiers        & CL             & getEpochCommittees                                                                                                                                                          \\
\midrule
peer\_id             & peer identifiers             & CL             & getPeers   \\
\bottomrule
\end{tabular}
}
\end{table}

\subsubsection{Fact-based Semantics-aware Request Generation.}
\label{subsubsec:mutation}
While syntactically valid requests conform to specification constraints, they may contain semantically invalid parameters that cause API calls to fail. For instance, a randomly generated block number may not exist yet, preventing clients from retrieving the requested data. To improve testing coverage, {\framework} employs a fact-based semantics-aware generation method that synthesizes requests with semantical meaningful values extracted from live blockchain data.

\noindent
\textbf{Fact Extraction.} As shown in Table~\ref{table:seman:mut}, {\framework} identifies 11 critical parameter types across Ethereum client APIs and maps them to 18 APIs that could be taken as valid data source. For example, for \texttt{block}, valid block numbers and hashes can be obtained from \texttt{eth\_getBlockReceipts}, while real addresses can be extracted from transaction data via \texttt{eth\_getTransactionReceipt}. This approach could transform the syntactically valid but semantically invalid request in Fig.~\ref{fig:eth_getBalance_valid} into the semantically valid request in Fig.~\ref{fig:eth_getBalance_sem_valid} by replacing the corresponding field.

\noindent
\textbf{Semantics Mutation.} Besides using extracted values directly, {\framework} generates additional semantically correct parameters through type-based mutation. 
Specifically, for parameter type defined in Table~\ref{table:seman:mut}, {\framework} applies tailored mutation operators relative to the current blockchain state.
Instead of generating arbitrary values, our strategy involves fetching the latest valid state from the target node and selecting historically valid values as test inputs.
For example, for CL client APIs requiring slot identifiers (time units for block proposals~\cite{slot}), {\framework} queries \texttt{getBlockHeader} to get the current slot and produces valid inputs by randomly selecting previous slot values $[1,\text{current\_slot}]$.
This method upholds semantic validity of the request while enhancing test coverage through increased parameter diversity.

\subsection{Phase II: Differential Testing}
\label{subsec:diff_testing}
After generating test requests, {\framework} conducts differential testing by deploying a controlled local testnet and comparing API responses across all client implementations.
This approach enhances test coverage while avoiding production network interference and the excessive costs associated with mainnet operations~\cite{run_node}.

Specifically, {\framework} deploys a local Ethereum testnet configured to mirror the latest mainnet version (Pectra~\cite{history}) to detect real-world bugs present in current implementations. The testnet includes 30 nodes covering all widely-adopted Ethereum clients: five EL clients and six CL clients that collectively represent 100\% and 99.93\% of mainnet deployments, respectively~\cite{diversity}. Each of the nodes in our testnet combines one EL client with one CL client, running in Docker containers and communicating via standardized APIs~\cite{engine}.
To minimize environmental false positives and ensure reliable comparisons (\textbf{Challenge \#2}), {\framework} standardizes the testing environment by configuring identical validator counts across all nodes and synchronizing the network state. Testing begins only after all nodes achieve full connectivity, reach the same block height, and the chain finalizes at least five epochs, approximately 40 minutes, to ensure network stability.

During testing execution, {\framework} randomly selects test requests and dispatches identical requests to all 30 nodes simultaneously. For each request, the system determines the appropriate HTTP method (\texttt{GET} or \texttt{POST}) and records comprehensive response data including the response body, HTTP status codes, and node identifiers. Any discrepancies observed across client responses are flagged as inconsistencies and forwarded to the next phase for further analysis.

\subsection{Phase III: Specification-Aware False Positive Filtering}
\label{subsec:bug}
When differential testing reveals inconsistencies across client responses, {\framework} must distinguish between genuine bugs and expected variations. Given the complexity of Ethereum API specifications, manual analysis would be inefficient. 
Thus, this phase employs a systematic false positive filtering approach that leverages API specifications and LLMs to identify real bugs. 

Large language models are responsible for understanding \textit{which fields should behave consistently across clients}, whose practicability has been proven by recent work~\cite{utfix, GPTScan, opdiffer, Artemis, test_expert}.
To achieve this, {\framework} utilizes a reasoning-integrated prompt that guides the LLM to perform semantic inference based on the specification.
Specifically, the prompt directs the LLM by (1) traversing the \texttt{result} field from API specifications (like Line 20 to Line 24 in Fig.~\ref{fig:spec:api}), and (2) performing a chain-of-thought~\cite{wei2022chain} analysis to infer the appropriate consistency policy based on the API summary and field descriptions (like Lines 3 and 17 in Fig.~\ref{fig:spec:api}). 
The LLM is asked to evaluate whether a field represents deterministic blockchain consensus state (labeled as \textit{must-identical}), local node metadata (\textit{may-divergent}), or unique instance identity (\textit{must-divergent}). 
For example, the \texttt{Balance} parameter is tagged as \textit{must-identical} (Line 26 of Fig.~\ref{fig:spec:api}) since account balances are determined by protocol and must be consistent across all clients. {\framework} then utilizes LLMs again, whose prompt is shown in Fig.~\ref{fig:checkprompt}, to reduce the three types of false positives identified in \textbf{Challenge \#2}.

\begin{itemize}[leftmargin=*]
    \item \textbf{Environmental State Differences (FP\#1):} Despite efforts to standardize the testing environment, blockchain growth and consensus protocol execution can cause legitimate node state variations. When inconsistencies involve fields labeled as \textit{may-divergent} and related to environmental or node-specific states, they are excluded as tolerable behaviors rather than bugs.
    \item \textbf{Specification-Allowed Variations (FP\#2):} The API specifications implicitly permit certain fields to differ across implementations. Inconsistencies in fields labeled as \textit{must-divergent} are treated as compliant behavior and filtered out to prevent false bug reports.
    \item \textbf{Semantic Equivalence (FP\#3):} The most challenging false positives occur when responses are semantically same yet syntactically different, such as representing the same numerical value in different formats. For fields labeled as \textit{must-identical} or \textit{may-divergent} that exhibit inconsistencies, {\framework} employs an LLM to check semantic  equivalence by analyzing both the API specifications and the divergent responses.
    As shown in the prompt in Fig.~\ref{fig:checkprompt}, the LLM assesses whether the observed discrepancies are genuine inconsistencies or semantically equivalent variations.
\end{itemize}
After applying these filtering rules, remaining inconsistencies are classified as genuine implementation bugs and reported to developer for investigation.

\begin{figure}
    \prompt{Prompt}{
        \textbf{System: }You are an expert on Ethereum execution layer client and Ethereum consensus layer client API specifications. \\
        \textbf{Task: }Evaluate the semantic equivalence of responses from multiple Ethereum clients to a specific API request, using the provided JSON schema as the ground truth. You will be given (1) a JSON schema describing the expected response structure and field semantics for a given API, (2) JSON responses from multiple clients for the same API request.\\
        \textbf{Instruction:} 
        \begin{enumerate}[label=\arabic*., leftmargin=*]
            \item For each response field marked as \textit{must-identical} or \textit{may-divergent}, compare its values across all client responses.
            \item If the field values differ syntactically, assess whether they are semantically equivalent based on specification.
            \item Report semantic inconsistency only when differing values violate meaning or behavior defined in the API response specification.
        \end{enumerate}
        \textbf{Input format:} (1) Target API JSON Schema (2) List of Client JSON Responses.\\
        \textbf{Output format:} Strictly return the following JSON object: \\
        {\ttfamily
        \{ \\
          "semantically\_equivalent": true/false, \\
          "reason": "<Concise technical justification>" \\
        \}} \\
        \textbf{Note:} No additional commentary outside the JSON object.
    }
    \vspace{-0.1in}
    \caption{An example prompt used to filter out false positives caused by semantically equivalent responses with syntactic differences.}
    \Description{An example prompt used to filter out false positives caused by semantically equivalent responses with syntactic differences.}
    \vspace{-0.2in}
    \label{fig:checkprompt}
\end{figure}
\section{Implementation \& Evaluation}

\noindent\textbf{Implementation.} We implement {\framework} in Python 3. 
The specifications used in the testing are JSON-RPC API~\cite{exec_api} and Beacon-API~\cite{beacon_api}.
To build the local Ethereum testnet, we leverage kurtosis~\cite{kurtosis} and ethereum-package~\cite{ether-package}.
We integrate OpenAI's GPT-5 model as the LLM and configure it with a temperature parameter of zero to get deterministic outputs.
We use \texttt{tx\_fuzz}~\cite{tx_fuzz} to send randomly generated transactions in the testnet. 

\noindent\textbf{Baseline Selection.}
To evaluate the effectiveness of {\framework}, we firstly consider domain-specific Ethereum client API testing tools. Following a comprehensive survey of the Ethereum client API testing landscape, we identified two baselines: EtherDiffer~\cite{kim2023etherdiffer} and \texttt{rpctestgen}~\cite{rpctg}.
EtherDiffer adopts a DSL-based approach to generate test inputs for EL client APIs, while \texttt{rpctestgen} is the official test generator maintained by Ethereum developers, relying on manually crafted EL client API test cases. As \texttt{rpctestgen} is limited to test generation, we utilize its output as input for execution within {\framework}.
We also consider state-of-the-art general-purpose RESTful API testing frameworks, where we select EvoMaster~\cite{evomaster,evomaster-github} as the baseline for CL client API testing.
EvoMaster is a popular search-based web API testing framework that employs evolutionary algorithms to generate test cases from API specifications. To enable EvoMaster to test CL client APIs, we manually modify the specification to remove certain fields, allowing it to read the spec and generate test cases.
In our evaluation, we utilize the latest stable releases of each tool, \textit{i.e.,} commit c8c13a3 for EtherDiffer~\cite{EtherDiffer-public}, commit 77b37aa for \texttt{rpctestgen}, and EvoMaster v4.0.0.

\noindent\textbf{Evaluation Targets.}
To systematically test the Ethereum client API implementations, we include all actively maintained EL and CL clients on the mainnet. Specifically, our evaluation covers five EL clients and six CL clients. These clients account for 100\% and 99.93\% of real-world deployments, respectively. 
Table~\ref{table:target:clients} in \S\ref{section:background:diversity} shows the details of the target clients.
Consequently, 30 Ethereum nodes are deployed in the testnet for API testing.

\noindent\textbf{Experiment Setup.}
All experiments were conducted on a desktop running Ubuntu 20.04, with Intel i7-12700 CPU and 128GB of RAM. In total,
we performed 14 rounds of testing on 11 Ethereum clients over an eight-month period (March to October 2025). In each round, we applied 20 test cases per API (5 syntactic-invalid, 5 syntactic-valid, and 10 semantic-valid). Experimental data illustrates that each round required an average of 40 minutes for environment setup, 45 minutes for test case generation, and 15 minutes for execution and inconsistency checks. Then we collected and reported all detected bugs to corresponding teams. Once they fixed bugs and pushed new client versions, we repeated the above testing process for a new round.

In summary, our experiments aim to answer the following research questions (RQs):
\begin{itemize}[leftmargin=2.5em]
    \item[\textbf{RQ1}] How do the code coverage and memory overhead of {\framework} compare to the baselines in terms of effectiveness and efficiency?
    \item[\textbf{RQ2}] How many real-world bugs in Ethereum client API implementations can be identified?
    \item[\textbf{RQ3}] What are the characteristics of the detected bugs? 
    \item[\textbf{RQ4}] How do the key components of {\framework} contribute to its performance?
\end{itemize}

To answer RQ1, we compare the code coverage and memory overhead of {\framework} with baselines. 
For RQ2, we present all the bugs identified by {\framework} and categorize these bugs according to the CWE system.
For RQ3, we conduct case studies and present developers response to reveal the real-world impact of detected bugs. 
For RQ4, we perform ablation studies to evaluate the contribution of each component of {\framework}.

\subsection{RQ1: Code Coverage \& Memory Usage}
To evaluate the effectiveness and efficiency of {\framework}, we assess both the code coverage of API implementations and the memory overhead in EL and CL clients.
Since EtherDiffer and \texttt{rpctestgen} only support EL client APIs and cannot be extended to CL clients due to limitations discussed in \S\ref{subsec:challenge}, we compare their code coverage for EL client APIs.
For CL client APIs, we compare the coverage results of {\framework} with those of EvoMaster.
As evaluating the code coverage requires instrumentation, we select Geth and Lighthouse, the most widely used EL and CL clients on the Ethereum mainnet~\cite{diversity} as the representatives.
For Geth, we measure the statement-level coverage of the entire client implementations, and API-specific packages. We use the official Go coverage tool~\cite{gocov} to perform the instrumentation. 
As for Lighthouse, we measure both function-level and statement-level coverage information by instrumenting with the LLVM instrumentation tools \texttt{llvm-profdata} and \texttt{llvm-cov}~\cite{rustcov}.

\begin{table}[t]
\caption{Code coverage and memory usage comparison between {\framework} and baselines. N/A indicates unsupported CL client APIs. \texttt{ethapi} and \texttt{catalyst} are API-related packages in Geth. F/S denote function/statement-level coverage for API packages (Pkg) and total codebase (Total). Mean and Peak represent the average and peak memory footprint during testing, respectively.}
\label{table:covage:results}
\resizebox{\columnwidth}{!}{%
\begin{tabular}{c|ccc|cccc|cc}
\toprule
    & \multicolumn{3}{c|}{\textbf{Geth (EL client)}}               & \multicolumn{4}{c|}{\textbf{Lighthouse (CL client)}} & \multicolumn{2}{c}{\textbf{Memory (MB)}}                                      \\
\multirow{-2}{*}{\textbf{Tool}}    & \textbf{\texttt{ethapi}} & \textbf{\texttt{catalyst}} & \textbf{Total} & \textbf{F (Pkg)} & \textbf{S (Pkg)} & \textbf{F (Total)} & \textbf{S (Total)} & \textbf{Mean} & \textbf{Peak} \\
\midrule

\rowcolor{codegray}
\textbf{\framework} & 43.20\%         & 40.40\%           & 32.40\%        & 57.08\%       & 65.83\%        & 17.95\%              & 19.14\%              & 41.84 & 70.16 \\
\textbf{EtherDiffer}               & 24.30\%         & 22.80\%           & 29.50\%        & N/A           & N/A            & N/A                  & N/A                  & 4,284.61 & 7,518.53\\
\textbf{rpctestgen}                & 34.20\%         & 21.30\%           & 29.00\%        & N/A           & N/A            & N/A                  & N/A   & 2.17 & 2.17 \\
\textbf{EvoMaster}                &  N/A          &  N/A            &  N/A         & 42.70\%           & 54.66\%           & 17.66\%                 & 18.82\% & 2,586.35 & 3,674.59 \\
\bottomrule
\end{tabular}
}
\vspace{-0.1in}
\end{table}

\textit{Code coverage.} As shown in Table~\ref{table:covage:results}, when evaluating the code coverage on entire client implementation (4th column), {\framework} achieves a code coverage of 32.40\%, outperforming the baselines by 9.83\% and 11.72\%, respectively.
We also conduct a fine-grained comparison in package level, \textit{i.e.,} focusing on \texttt{ethapi} and \texttt{catalyst} packages, which are specifically designed for handling APIs.
Specifically, on these two packages, {\framework} achieves code coverage of 43.20\% and 40.40\%, representing improvements of up to 77.77\% and 89.67\% over the state-of-the-art baselines, respectively. 
For CL client API implementations, {\framework} achieves code coverage of 65.83\% at the statement level and 57.08\% at the function level for API-specific packages. These figures reflect increases of 20.44\% and 33.68\% over the baseline. Similarly, at the overall codebase level, {\framework} reaches 19.14\% statement-level coverage and 17.95\% function-level coverage, with improvements of 1.70\% and 1.64\% over the baseline. Notably, while {\framework} demonstrates substantial coverage improvements in API-specific packages, the increases at the codebase level are relatively incremental. This is primarily because API logic represents only a limited subset of the whole codebase. 
Consequently, coverage gains within API-related modules are inevitably marginalized by the extensive volume of non-API code.
This confirms that our approach effectively targets and exercises the intended API surface, leading to global increments that are consistent with the specialized scope of API fuzzing.

After a thorough review, we summary two factors that contribute to the coverage improvement of {\framework} over the baselines:
\begin{itemize}[leftmargin=*]
    \item \textbf{Automated and systematic request generation capabilities.} Our approach could systematically generates test requests for all Ethereum client APIs defined in specifications. In contrast, existing DSL-based or template-based manually written test cases are often outdated and fail to fully test the latest API implementations. Moreover, manual composing processes are tedious and error-prone.
    \item \textbf{Exploration of corner cases.} Our method deliberately introduces  syntactically invalid and semantically invalid test inputs, which are often overlooked by manual composing methods but could trigger unusual corner cases~\cite{geth_corner}, thereby increasing the coverage of edge scenarios.
\end{itemize}

\textit{Memory usage.} To evaluate the efficiency of {\framework}, we compare its memory overhead with baselines. We consider both the average memory usage and peak memory usage. As illustrated in Table~\ref{table:covage:results}, {\framework} achieves an average memory usage of 41.84 MB and and a peak of 70.16 MB. 
Regarding the baselines, while \texttt{rpctestgen} exhibits the lowest memory usage of 2.17 MB, it should be noted that \texttt{rpctestgen} is a test generator and does not perform large-scale testing.
In contrast, the memory usage of EtherDiffer is 100 times of {\framework} because besides testing targets, it runs another 12 Geth clients for test case generation.
Similarly, the memory usage of EvoMaster is still in the GB level and may throw out-of-memory errors, as it runs on JVM and employs a search-based method for test generation, which can be memory-intensive. 

We intentionally exclude the memory usage of the local testnet from our measurements to provide a fair assessment of the {\framework}'s intrinsic overhead. Specifically, the testnet serves as our testing target and its memory usage is approximately 15 GB. To maximize the capability of detecting real-world bugs, we deploy 30 nodes running production Ethereum clients.  Consequently, including this memory usage would overshadow the actual efficiency of {\framework}'s testing logic.

\answerbox{\textbf{\noindent Answer for RQ1: } {Compared with three state-of-the-art baselines, in terms of coverage, {\framework} achieves up to 11.72\% improvement on whole EL client codebases and 89.67\% improvement in API-specific packages. {\framework} also demonstrates its ability to handle CL client APIs. For memory, achieves an average memory usage of 41.84 MB and and a peak of 70.16 MB. They jointly prove the effectiveness and efficiency of {\framework}.
}}

\begin{table}[t]
\caption{Classification of detected bugs by CWE IDs.}
\vspace{-0.1in}
\label{table:cwe}
\resizebox{1\columnwidth}{!}{%
\begin{tabular}{lcl}
\toprule
\multicolumn{1}{c}{\textbf{CWE ID}}                                     & \textbf{\#} & \multicolumn{1}{c}{\textbf{Bug IDs}}                                                                                                         \\
\midrule

\multicolumn{1}{l}{CWE-684: Incorrect Provision of   Specified Functionality}     & 33          & \begin{tabular}[c]{@{}l@{}} 1–4, 6, 9–10, 12, 15, 16-18, 22-29, 31-33, \\ 41-43, 45, 51, 55, 58, 63, 65, 72        \end{tabular} \\
 
CWE-755: Improper Handling of   Exceptional Conditions                            & 10           &  8, 13, 14, 44, 50, 57,   61, 62 ,64 ,66                                                                                                                     \\

CWE-704: Incorrect Type Conversion or   Cast                                      & 8           &  19, 40, 46, 53, 54, 60 ,67 ,71                                                                                                               \\

CWE-1287: Improper Validation of   Specified Type of Input                        & 5           &  5, 20-21, 39, 52                                                                                                                                      \\

CWE-393: Return of Wrong Status Code                                              & 3           &  36, 38, 70                                                                                                                                        \\

\multicolumn{1}{l}{\begin{tabular}[c]{@{}l@{}}CWE-703: Improper Check or Handling of Exceptional \\ \hspace*{4.2em} Conditions\end{tabular}} & 3 &  7, 34, 48\\

CWE-325: Missing Cryptographic Step                                               & 2           &  11, 30                                                                                                                                         \\

CWE-544: Missing Standardized Error Handling Mechanism                          & 2           &   56                                                                                                                                       \\

\multicolumn{1}{l}{\begin{tabular}[c]{@{}l@{}}CWE-758: Reliance on Undefined, Unspecified, or \\ \hspace*{4.4em}Implementation-Defined Behavior\end{tabular}} & 2 & 47, 68 \\
 
CWE-121: Stack-based Buffer Overflow                                              & 2           &    49, 69                                                                                                                                       \\

CWE-116: Improper Encoding or Escaping   of Output                                & 1           &      37                                                                                                                                        \\
 
CWE-390: Detection of Error Condition   Without Action                            & 1           &    59                                                                                                                                          \\
 
CWE-697: Incorrect Comparison                                                     & 1           &      35               \\              \bottomrule                                                   \end{tabular}
}
\vspace{-0.2in}
\end{table}

\subsection{RQ2: Real-World API Bugs}
Beyond code coverage, we evaluate the effectiveness of {\framework} in terms of its ability to identify real-world Ethereum client API bugs.
In total, {\framework} identified 72 bugs, \textbf{87.50\%} (63/72) of which were firstly uncovered by us, as illustrated in Table~\ref{table:real:world:bugs:short}.
Our testing reveals that \textit{\textbf{Ethereum client API bugs are widespread in the whole Ethereum ecosystem, \textit{i.e.,} no mainstream EL and CL clients are spared.}}
Since the testing process of {\framework} can generate specific requests for all bugs, which can be easily reproduced and confirmed by developers, most responses from developers are relatively positive.
Specifically, a total of 65 bugs have been confirmed by developers, with only 7 bugs still awaiting response at the time of paper submission. For confirmed bugs, 81.54\% (53/65) of bugs have been fixed already, 7.69\% (5/65) of bugs are scheduled to be fixed, and 10.77\% (7/65) of bugs are marked as ``Won't fix'' due to the API being deprecated or the corresponding inconsistency is expected.

\clearpage
\begin{landscape}
\begin{table}[p]
\caption{Details of 72 bugs identified by {\framework} across all Ethereum clients, where \scalebox{0.7}{\faIcon{check}} and \scalebox{0.7}{\faIcon{times}} of the \textbf{Unk.} column indicate the bug is unknown or known, respectively. \textbf{Sta.} denotes status, where \scalebox{0.7}{\faIcon{circle}}, \scalebox{0.7}{{\faIcon{adjust}}},  \scalebox{0.7}{\faIcon[regular]{circle}}, and \scalebox{0.7}{\faIcon[regular]{dot-circle}} represent the bug is fixed, will fix, won't fix or reported,  respectively.}
\label{table:real:world:bugs:short}
\resizebox{\linewidth}{!}{%
\begin{tabular}{c|c|cccc|c|c|ccccc}
\toprule
\multicolumn{1}{c|}{\textbf{No.}} & \multicolumn{1}{c|}{\textbf{Type}} & \textbf{Project}     & \textbf{Unk.}                                                      & \textbf{Sta.}                                                                        & \multicolumn{1}{c|}{\textbf{Root cause}}                    & \multicolumn{1}{c|}{\textbf{No.}} & \multicolumn{1}{c|}{\textbf{Type}} & \textbf{Project}          & \textbf{Unk.}                                                      & \textbf{Sta.}                                                                        & \textbf{Root cause}                                                &                      \\ \cline{1-12}
\multicolumn{1}{c|}{1}            & \multirow{36}{*}{CL Client}     & \multirow{3}{*}{Lighthouse}           & \scalebox{0.7}{\faIcon{check}} & \scalebox{0.7}{\faIcon{circle}}                  & \multicolumn{1}{c|}{Missing field in the API response}      & \multicolumn{1}{c|}{37}           & \multirow{8}{*}{CL Client}     & \multirow{4}{*}{Lodestar} & \scalebox{0.7}{\faIcon{times}} & \scalebox{0.7}{\faIcon{circle}}                  & Incorrect encoding in API responses                                &                      \\
\multicolumn{1}{c|}{2}            &              &                      & \scalebox{0.7}{\faIcon{check}} & \scalebox{0.7}{\faIcon{circle}}                  & \multicolumn{1}{c|}{Missing field in the API response}      & \multicolumn{1}{c|}{38}           &              &                           & \scalebox{0.7}{\faIcon{times}} & \scalebox{0.7}{\faIcon{circle}}                  & Error handling for invalid parameter                               &                      \\
\multicolumn{1}{c|}{3}            &              &                      & \scalebox{0.7}{\faIcon{check}} & \scalebox{0.7}{\faIcon{circle}}                  & \multicolumn{1}{c|}{Missing API implementation in the spec} & \multicolumn{1}{c|}{39}           &              &                           & \scalebox{0.7}{\faIcon{check}} & \scalebox{0.7}{\faIcon{circle}}                  & API fails to parse input parameter                                 &                      \\
\cline{3-6}
4                                 &                                    & \multirow{13}{*}{Prysm}                & \scalebox{0.7}{\faIcon{times}} & \scalebox{0.7}{\faIcon{circle}}                  & Missing API implementation in the spec                      & 40                                &                                    &                           & \scalebox{0.7}{\faIcon{check}} & \scalebox{0.7}{\faIcon{circle}}                  & API use wrong data type from old fork                              &                      \\ \cline{9-12}
\multicolumn{1}{c|}{5}            &              &                      & \scalebox{0.7}{\faIcon{check}} & \scalebox{0.7}{\faIcon{circle}}                  & \multicolumn{1}{c|}{API fails to parse input parameter}     & \multicolumn{1}{c|}{41}           &              & \multirow{4}{*}{Grandine}                  & \scalebox{0.7}{\faIcon{check}} & \scalebox{0.7}{\faIcon{circle}}                  & Missing API implementation in the spec                             &                      \\
\multicolumn{1}{c|}{6}            &              &                      & \scalebox{0.7}{\faIcon{check}} & \scalebox{0.7}{\faIcon{circle}}                  & \multicolumn{1}{c|}{Missing API implementation in the spec} & \multicolumn{1}{c|}{42}           &              &                           & \scalebox{0.7}{\faIcon{check}} & \scalebox{0.7}{\faIcon{circle}}                  & Missing API implementation in the spec                             &                      \\
\multicolumn{1}{c|}{7}            &              &                      & \scalebox{0.7}{\faIcon{check}} & \scalebox{0.7}{\faIcon{circle}}                  & \multicolumn{1}{c|}{Error handling for empty slashing}      & \multicolumn{1}{c|}{43}           &              &                           & \scalebox{0.7}{\faIcon{check}} & \scalebox{0.7}{\faIcon{circle}} & Missing API implementation in the spec                             &                      \\
8                                 &                                    &                      & \scalebox{0.7}{\faIcon{check}} & \scalebox{0.7}{\faIcon{circle}}                  & Error handling for invalid parameter                        & 44                                &                &       & \scalebox{0.7}{\faIcon{check}} & \scalebox{0.7}{\faIcon[regular]{dot-circle}} & Error handling for invalid parameter                               &                      \\ \cline{8-12}
\multicolumn{1}{c|}{9}            &              &                      & \scalebox{0.7}{\faIcon{check}} & \scalebox{0.7}{\faIcon{circle}}                  & \multicolumn{1}{c|}{Missing field in the API response}      & \multicolumn{1}{c|}{45}           & \multirow{25}{*}{EL Client}     & \multirow{4}{*}{Geth}                     & \scalebox{0.7}{\faIcon{check}} & \scalebox{0.7}{\faIcon{circle}}                  & Extra field in the API response                                    &                      \\
\multicolumn{1}{c|}{10}           &              &                      & \scalebox{0.7}{\faIcon{check}} & \scalebox{0.7}{\faIcon{circle}}                  & \multicolumn{1}{c|}{Incorrect field in the API response}    & \multicolumn{1}{c|}{46}           &              &                           & \scalebox{0.7}{\faIcon{check}} & \scalebox{0.7}{\faIcon[regular]{circle}}     & Handling of edge cases in API response values                      &                      \\
\multicolumn{1}{c|}{11}           &              &                      & \scalebox{0.7}{\faIcon{check}} & \scalebox{0.7}{\faIcon{circle}}                  & \multicolumn{1}{c|}{Missing signature validation in API}    & \multicolumn{1}{c|}{47}           &              &                           & \scalebox{0.7}{\faIcon{check}} & \scalebox{0.7}{\faIcon[regular]{circle}}     & Inconsistent support for deprecated API                            &                      \\
12                                &                                    &                      & \scalebox{0.7}{\faIcon{times}} & \scalebox{0.7}{\faIcon{circle}}                  & Missing API implementation in the spec                      & 48                                &                &       & \scalebox{0.7}{\faIcon{check}} & \scalebox{0.7}{\faIcon{circle}}                  & \multicolumn{1}{c}{Handling of edge cases in API parameter}        &                      \\ \cline{9-12}
\multicolumn{1}{c|}{13}           &              &                      & \scalebox{0.7}{\faIcon{check}} & \scalebox{0.7}{\faIcon{circle}}                  & \multicolumn{1}{c|}{Error handling for invalid parameter}   & \multicolumn{1}{c|}{49}           &              & \multirow{6}{*}{Nethermind}                & \scalebox{0.7}{\faIcon{check}} & \scalebox{0.7}{\faIcon{adjust}}                  & Stack overflow during evm execution                                &                      \\
\multicolumn{1}{c|}{14}           &              &                      & \scalebox{0.7}{\faIcon{check}} & \scalebox{0.7}{\faIcon{circle}}                  & \multicolumn{1}{c|}{Incorrect initialization of the chain}  & \multicolumn{1}{c|}{50}           &              &                           & \scalebox{0.7}{\faIcon{check}} & \scalebox{0.7}{\faIcon{circle}}                  & Error handling for non-existent blocks                             &                      \\
\multicolumn{1}{c|}{15}           &              &                      & \scalebox{0.7}{\faIcon{check}} & \scalebox{0.7}{\faIcon{circle}}                  & \multicolumn{1}{c|}{Valid parameters are not supported}     & \multicolumn{1}{c|}{51}           &              &                           & \scalebox{0.7}{\faIcon{times}} & \scalebox{0.7}{\faIcon[regular]{dot-circle}} & Deprecated field in the API response                               &                      \\

\cline{3-6}

16                                &                                    &                      & \scalebox{0.7}{\faIcon{check}} & \scalebox{0.7}{\faIcon{circle}}                  & Error handling for invalid parameter                        & 52                                &                                    &                           & \scalebox{0.7}{\faIcon{check}} & \scalebox{0.7}{\faIcon{circle}}                  & Missing check for Transaction                                      &                      \\

17                                &                                    & \multirow{5}{*}{Teku}                 & \scalebox{0.7}{\faIcon{check}} & \scalebox{0.7}{\faIcon{circle}}                  & Missing API implementation in the spec                      & 53                                &                                    &                           & \scalebox{0.7}{\faIcon{check}} & \scalebox{0.7}{\faIcon{circle}}                  & Type conversion error of filter\_id                                &                      \\
18                                &                                    &                      & \scalebox{0.7}{\faIcon{times}} & \scalebox{0.7}{\faIcon{adjust}}                  & Missing API implementation in the spec                      & 54                                &                                    &                           & \scalebox{0.7}{\faIcon{check}} & \scalebox{0.7}{\faIcon{circle}}                  & Handling of edge cases in API response values                      &                      \\ \cline{9-12}
19                                &                                    &                      & \scalebox{0.7}{\faIcon{check}} & \scalebox{0.7}{\faIcon{adjust}}                  & Wrong type in the API response                              & 55                                &                                    & \multirow{4}{*}{Besu}                      & \scalebox{0.7}{\faIcon{times}} & \scalebox{0.7}{\faIcon{adjust}}                  & Deprecated field in the API response                               &                      \\
\multicolumn{1}{c|}{20}           &              &                      & \scalebox{0.7}{\faIcon{check}} & \scalebox{0.7}{\faIcon{circle}}                  & \multicolumn{1}{c|}{Fail to parse valid parameter}          & \multicolumn{1}{c|}{56}           &              &                           & \scalebox{0.7}{\faIcon{check}} & \scalebox{0.7}{\faIcon[regular]{circle}}     & Error handling for invalid parameter                               &                      \\
\multicolumn{1}{c|}{21}           &              &                      & \scalebox{0.7}{\faIcon{check}} & \scalebox{0.7}{\faIcon{circle}}                  & \multicolumn{1}{c|}{Schema definition error}                & \multicolumn{1}{c|}{57}           &              &                           & \scalebox{0.7}{\faIcon{check}} & \scalebox{0.7}{\faIcon{circle}}                  & Error handling for invalid parameter                               &                      \\
\cline{3-6}
\multicolumn{1}{c|}{22}           &              & \multirow{9}{*}{Nimbus}          & \scalebox{0.7}{\faIcon{check}} & \scalebox{0.7}{\faIcon{circle}}                  & \multicolumn{1}{c|}{Missing field in the API response}      & \multicolumn{1}{c|}{58}           &              &                           & \scalebox{0.7}{\faIcon{check}} & \scalebox{0.7}{\faIcon[regular]{circle}}     & Handling of edge cases in API response values                      &                      \\ \cline{9-12}
\multicolumn{1}{c|}{23}           &              &                      & \scalebox{0.7}{\faIcon{check}} & \scalebox{0.7}{\faIcon{circle}}                  & \multicolumn{1}{c|}{Missing field in the API response}      & \multicolumn{1}{c|}{59}           &              & \multirow{6}{*}{Erigon}                    & \scalebox{0.7}{\faIcon{check}} & \scalebox{0.7}{\faIcon{circle}}                  & Method handler crashed for nil                                     &                      \\
24                                &                                    &                      & \scalebox{0.7}{\faIcon{check}} & \scalebox{0.7}{\faIcon{circle}}                  & Missing field in the API response                           & 60                                &                                    &                           & \scalebox{0.7}{\faIcon{check}} & \scalebox{0.7}{\faIcon{circle}}     & Handling of edge cases in API response values                      &  \\
25                                &                &  & \scalebox{0.7}{\faIcon{check}} & \scalebox{0.7}{\faIcon{circle}}                  & \multicolumn{1}{c|}{Missing API implementation in the spec}  & 61                                &                                    &                           & \scalebox{0.7}{\faIcon{check}} & \scalebox{0.7}{\faIcon{circle}}                  & Error handling for invalid ids                                     &                      \\ 
\multicolumn{1}{c|}{26}           &              &                      & \scalebox{0.7}{\faIcon{check}} & \scalebox{0.7}{\faIcon{adjust}}     & \multicolumn{1}{c|}{Missing API implementation in the spec} & \multicolumn{1}{c|}{62}           &              &                           & \scalebox{0.7}{\faIcon{check}} & \scalebox{0.7}{\faIcon{circle}}                  & Error handling for invalid ids                                     &  \\ 
\multicolumn{1}{c|}{27}           &              &                      & \scalebox{0.7}{\faIcon{check}} & \scalebox{0.7}{\faIcon[regular]{circle}}     & \multicolumn{1}{c|}{Extra field in the API response}        & \multicolumn{1}{c|}{63}           &              &                           & \scalebox{0.7}{\faIcon{check}} & \scalebox{0.7}{\faIcon{circle}}                  & Extra field in the API response                                    &  \\
\multicolumn{1}{c|}{28}           &              &                      & \scalebox{0.7}{\faIcon{check}} & \scalebox{0.7}{\faIcon{circle}}                  & \multicolumn{1}{c|}{Missing field in the API response}      & \multicolumn{1}{c|}{64}           &              &                           & \scalebox{0.7}{\faIcon{check}} & \scalebox{0.7}{\faIcon{circle}}                  & Error handling for empty parameter                                 &  \\ \cline{9-12}
\multicolumn{1}{c|}{29}           &              &                      & \scalebox{0.7}{\faIcon{check}} & \scalebox{0.7}{\faIcon[regular]{dot-circle}}     & \multicolumn{1}{c|}{Wrong data type of the API response}    & \multicolumn{1}{c|}{65}           &              & \multirow{5}{*}{Reth}                     & \scalebox{0.7}{\faIcon{check}} & \scalebox{0.7}{\faIcon{circle}}                  & Missing API implementation in the spec                             &                      \\
\multicolumn{1}{c|}{30}           &              &                      & \scalebox{0.7}{\faIcon{check}} & \scalebox{0.7}{\faIcon[regular]{dot-circle}} & \multicolumn{1}{c|}{Missing signature validation in API}    & \multicolumn{1}{c|}{66}           &              &                           & \scalebox{0.7}{\faIcon{times}} & \scalebox{0.7}{\faIcon{circle}}                  & Error handling for empty block                                     &                      \\
\cline{3-6}
\multicolumn{1}{c|}{31}           &              & \multirow{6}{*}{Lodestar}             & \scalebox{0.7}{\faIcon{check}} & \scalebox{0.7}{\faIcon{circle}}                  & \multicolumn{1}{c|}{Missing field in the API response}      & \multicolumn{1}{c|}{67}           &              &                           & \scalebox{0.7}{\faIcon{check}} & \scalebox{0.7}{\faIcon[regular]{circle}}     & Handling of edge cases in API response values                      &                      \\
32                                &                                    &                      & \scalebox{0.7}{\faIcon{check}} & \scalebox{0.7}{\faIcon{circle}}                  & Empty field in the API response                             & 68                                &                                    &                           & \scalebox{0.7}{\faIcon{check}} & \scalebox{0.7}{\faIcon[regular]{circle}}     & Inconsistent support for deprecated API                            &                      \\
\multicolumn{1}{c|}{33}           &              &                      & \scalebox{0.7}{\faIcon{check}} & \scalebox{0.7}{\faIcon{circle}}                  & \multicolumn{1}{c|}{Extra field in the API response}        & \multicolumn{1}{c|}{69}           &              &       & \scalebox{0.7}{\faIcon{check}} & \scalebox{0.7}{\faIcon[regular]{dot-circle}}                  & \multicolumn{1}{c}{Stack underflow of missing parameter check} &                      \\ \cline{9-12}
\multicolumn{1}{c|}{34}           &              &  & \scalebox{0.7}{\faIcon{check}} & \scalebox{0.7}{\faIcon{circle}}                  & \multicolumn{1}{c|}{Wrong type in the API request}          & \multicolumn{1}{c|}{70}           & \multirow{3}{*}{Spec}           &  \multirow{2}{*}{JSON-RPC API}              & \scalebox{0.7}{\faIcon{check}} & \scalebox{0.7}{\faIcon[regular]{dot-circle}} & Call to standardize error handling logic                           &                      \\
\multicolumn{1}{c|}{35}           &              &                      & \scalebox{0.7}{\faIcon{check}} & \scalebox{0.7}{\faIcon{circle}}                  & \multicolumn{1}{c|}{Extra validation in the API response}   & \multicolumn{1}{c|}{71}           &              &                           & \scalebox{0.7}{\faIcon{check}} & \scalebox{0.7}{\faIcon[regular]{dot-circle}} & Wrong value in the API request                                     &                      \\ \cline{9-12}
\multicolumn{1}{c|}{36}           &              &                      & \scalebox{0.7}{\faIcon{check}} & \scalebox{0.7}{\faIcon{circle}}                  & \multicolumn{1}{c|}{Error handling for invalid parameter}   & \multicolumn{1}{c|}{72}           &               & Beacon-API                & \scalebox{0.7}{\faIcon{times}} & \scalebox{0.7}{\faIcon{circle}}                  & Schema inconsistent with CL spec                                   &                     \\
\bottomrule
\end{tabular}
}
\end{table}
\end{landscape}

Interestingly, {\framework} also uncovers bugs in the Ethereum API specifications. 
Since test requests are generated based on the specification, any inconsistency between the generated request and the client's input requirements may indicate a flaw in the specification itself. For example, the specification defines the field as 32, while the implementation requires 33.
To determine whether an inconsistency stems from the specification or a client API implementation error, we conduct manual analysis by cross-checking the behavior across multiple clients and verifying against the intent stated in the specifications.
In total, we have identified three bugs in the Ethereum node API specifications. 
The \textit{case study \#1} in \S\ref{sec:case_study1} details the impact and detection method for one of the specification bugs.

To assess the distribution of bugs and their potential impact, we categorize identified bugs according to the well-known Common Weakness Enumeration (CWE)~\cite{cwe} based on the developers’ responses and independent evaluations from two authors.
As shown in Table~\ref{table:cwe}, the 72 bugs fall into 13 CWE categories. The majority (45.83\%) of bugs are classified into CWE-684, which indicates that the implementation does not conform to the specification, often due to missing fields or unimplemented APIs. 
Another significant portion of the bugs (19.44\%, 14 bugs) are related to improper exception handling, corresponding to CWE-755, CWE-703, and CWE-390.
In addition, there are also bugs related to stack overflows, input validation, and output encoding, indicating that the root causes and manifestations of API bugs are diverse.

The real-world bug identification provides several key insights:
\begin{itemize}[leftmargin=*]
    \item \textbf{Bugs are everywhere.} Due to the rapid upgrade cycles of Ethereum protocol and the collaborative nature of development across different teams, Ethereum client API bugs are widespread. This highlights the need for coordinated testing efforts involving multiple client teams.
    \item \textbf{Same bug, but different attitudes.} Developers from different teams may have inconsistent attitudes toward the same API bug. For example, regarding deprecated methods predating the PoS transition (\textit{e.g.,} \texttt{eth\_coinbase}), Geth and Reth have dropped support, whereas Besu, Nethermind and Erigon continue to maintain its compatibility.
    \item \textbf{API Bugs primarily stem from specification non-compliance.} Nearly half of the identified bugs (45.83\%) fall under CWE-684, indicating implementation-specification mismatches rather than traditional security flaws like buffer overflows or injection vulnerabilities. This suggests that API development faces a fundamental challenge in correctly interpreting and fully implementing complex specifications, with developers more likely to err in specification adherence than in basic secure coding practices.
\end{itemize}

\answerbox{\textbf{\noindent Answer for RQ2: } {\framework} has detected 72 bugs in all major Ethereum EL and CL clients, of which 90.28\% (65/72) have been confirmed or fixed. Interestingly, {\framework} also identified three bugs in official specifications, indicating its effectiveness in detecting real-world Ethereum client API bugs. These findings highlight the necessity and value of {\framework}.}

\subsection{RQ3: Bug Characterization}
To characterize identified bugs, we conduct case studies of two representative bugs.
Then, we share responses from client development teams to illustrate the real-world impact of the reported bugs.

\begin{figure}[t]
\centering
\begin{lstlisting}[language=myjson,breaklines=true,xleftmargin=1em,framexleftmargin=1em]
{
  "body": {
    "randao_reveal": "...",
    "...": "...",
    "deposits": {
      "proof": {
        "...": "...",
// Bug here: Both values should be 33
        "minItems": 32,
        "maxItems": 32
      },
      "data": "..."
    },
    "execution_payload": {
      "parent_hash": "...",
      "...": "...",
// Bug here: Missing definition
//    "blob_gas_used": "...",
      "block_hash": "..."
    }
  }
}
\end{lstlisting}
\vspace{-0.1in}
\caption{A specification bug in Beacon-API where the \texttt{proof} array size is incorrectly defined as 32 instead of 33 and the EIP-4844 \texttt{blob\_gas\_used} field is missing.}
\Description{A specification bug in Beacon-API where the \texttt{proof} array size is incorrectly defined as 32 instead of 33 and the EIP-4844 \texttt{blob\_gas\_used} field is missing.}
\label{fig:case_study1}
\end{figure}

\paragraph{Case Study \#1: Incorrect API Specification. }
\label{sec:case_study1}
The bug in Fig.~\ref{fig:case_study1}\footnote{For clarity, we simplify the request structure in Fig.~\ref{fig:case_study1}, while the full specification is available in~\cite{publishBlockV2}.} corresponds to No.72 in Table~\ref{table:real:world:bugs:short}, a bug identified in the Beacon-API specification.  
Specifically, the specification sets the required size for \texttt{proof} as 32, whereas the correct value defined in the consensus rules is actually 33, resulting in a discrepancy in the expected value.
The CL client API \texttt{publishBlockV2} includes the \texttt{blob\_gas\_used} field, introduced in EIP-4844~\cite{eip-4844} during the Ethereum Dencun upgrade~\cite{history}. Despite its implementation in March 2024, this field remains notably absent from the Beacon-API specification as of March 2025.
Given that CL clients implement APIs based on the specification and users typically refer to the specification to understand how to use the API, this bug could propagate to downstream implementations or negatively impact the Ethereum user experience.

{\framework} identified this bug through its ability to automatically generate requests based on the specification, as described in \S\ref{subsec:testcase_generate}. Specifically, we generate erroneous test requests based on the faulty specification, and during differential testing, the clients reject all the test requests. The error message from the clients ``expected 33 and 32 found''  indicates that the specification is incorrect. We further investigated this manually and found the root cause. Fortunately, the Beacon-API official confirmed this bug within 15 minutes of our report and promptly pushed a fix within an hour. 
Another developer from a different CL client team opened an issue to investigate whether this inconsistency also affects their API implementation.

\begin{figure}
    \centering
\begin{lstlisting}[
        language=go,
        style=gocomment,
        numbers=left,    
        numbersep=4pt,                   
        xleftmargin=1em,
        framexleftmargin=1em]
func (api *APIImpl) CreateAccessList(...) (...) {
    ...
    a, err := stateReader.ReadAccountData(*args.From)
    if err != nil {
        return nil, err
    }
!!++  if a == nil {
++      return nil, errors.New("Account: " + args.From.Hex() + " not found")
++  }!!
    nonce = a.Nonce + 1
    ...
}
\end{lstlisting}
    \vspace{-0.1in}
    \caption{An unhandled nil pointer exception in Erigon's \texttt{CreateAccessList} API and the corresponding fix, which adds proper account existence checks.}
    \Description{An unhandled nil pointer exception in Erigon's \texttt{CreateAccessList} API and the corresponding fix, which adds proper account existence checks.}
    \vspace{-0.1in}
    \label{fig:case_study2}
\end{figure}

\paragraph{Case Study \#2: Unhandled Exception Bug. }  
The bug in Fig.~\ref{fig:case_study2} corresponds to No.59 in Table~\ref{table:real:world:bugs:short}, a bug of EL client, Erigon~\cite{erigon}. 
This bug stems from a missing exception handler in the API \texttt{eth\_createAccessList}, which is responsible for generating an access list for a given transaction. 
Specifically, when the \texttt{from} address in the transaction refers to a non-existent account, the API attempts to access this account without a prior existence check. In Erigon, this results in a \texttt{nil} pointer dereference and triggers the error message ``method handler crashed''. 
{\framework} detected this bug via the generation of syntactically valid but semantically invalid test requests, which triggered a corner case where the exception was not handled, leading to a crash.
It was fixed by introducing proper error-handling logic to explicitly report the absence of the sender account.

\paragraph{Developers Response.}
We report the detected bugs primarily through GitHub issues. We also actively contribute fixes for identified API bugs to avoid potential exploitation from the adversary.
Our testing results received positive feedback from client developers, including acknowledgments of the bug reports, expressed interest in our testing approach, integration of unit test cases for the corner cases we discovered, and special acknowledgments in the client version release notes.

Here we quote several selected responses from client teams:

\begin{center}
\textit{``Thanks for raising the difference in behavior between EL clients.''} \\

\textit{``Really cool to see that more people start using the spec schema to test implementation.''}

\textit{``Special thanks to external contributors!''}\\

\textit{``I am curious though how you found them.''}
\end{center}

Notably, one bug identified in the EL client Besu, specifically in the \texttt{engin\_newPayloadV4} API, was brought up for discussion during the regular Ethereum Project Management meeting on RPC standards~\cite{pm_rpc}, reflecting the broader community’s recognition and interest.

\answerbox{\textbf{\noindent Answer for RQ3: } 
{\framework}'s findings demonstrate significant real-world impact through diverse bug types ranging from specification errors to unhandled exceptions, with developers providing positive feedback including acknowledgments, integration of test cases, and expressed interest in the testing methodology. Notably, one identified bug was escalated to the Ethereum Project Management meeting for discussion, highlighting the critical importance of these discoveries to the Ethereum ecosystem.

}

\subsection{RQ4: Ablation Studies}
To assess the contribution of the key components in {\framework}, considering whether the component could be removed without impacting the functional integrity, we decided to perform two ablation studies. 
Specifically, we measured the contribution of the \textit{fact-based semantics-aware generation (FSG)} (see \S\ref{subsubsec:mutation}) and \textit{false positive filter (FPF)} (see \S\ref{subsec:bug}).

\begin{table}[t]
\caption{Code coverage results when disabling semantics-aware request generation (FSG).}
\vspace{-0.1in}
\label{table:coverage:ablation}
\begin{tabular}{c|ccc|cccc}
\toprule
    & \multicolumn{3}{c|}{\textbf{Geth (EL client)}}               & \multicolumn{4}{c}{\textbf{Lighthouse (CL client)}}                                       \\
\multirow{-2}{*}{\textbf{Tool}}    & \textbf{\texttt{ethapi}} & \textbf{\texttt{catalyst}} & \textbf{Total} & \textbf{F (Pkg)} & \textbf{S (Pkg)} & \textbf{F (Total)} & \textbf{S (Total)} \\
\midrule
w/o FSG               & 35.61\%         & 34.18\%           & 29.50\%        & 45.27\%           &       52.21\%      & 15.13\%                  & 16.15\%                  \\
\rowcolor{codegray}
\textbf{\framework} & 43.20\%         & 40.40\%           & 32.40\%        & 57.08\%       & 65.83\%        & 17.95\%              & 19.14\%              \\
\bottomrule
\end{tabular}

\end{table}

On the one hand, as shown in Table~\ref{table:coverage:ablation}, the results demonstrate that FSG positively contributes to the code coverage during the testing process.
Quantitatively, when disabling FSG, on Geth, the statement-level coverage decreases 8.95\% and 17.57\% when considering the whole codebase and API-specific packages, respectively; on Lighthouse, the statement-level coverage drops by 15.62\% and 20.69\% when considering the whole codebase and API-specific packages, respectively.
We emphasize that the decrease in coverage is mainly due to the lack of FSG when generating API requests, which cannot guarantee the semantic validity, resulting in some seemingly reasonable but actually impossible API requests. For example, retrieving the block hash with a block height that is too high to exist.
FSG mitigates this issue by injecting valid semantic content retrieved from the chain through specific API calls (detailed in \S\ref{subsubsec:mutation}).

\begin{table}[t]
\vspace{-0.15in}
\caption{False discovery rate (FDR) comparison with and without the false positive filter (FPF) component.}
\vspace{-0.05in}
\label{table:FDR:ablation}
\begin{tabular}{c|cccc|cccc}
\toprule
    & \multicolumn{4}{c|}{\textbf{Geth (EL client)}}               & \multicolumn{4}{c}{\textbf{Lighthouse (CL client)}}                                       \\
\multirow{-2}{*}{\textbf{Tool}}    & \textbf{TP} & \textbf{FP} & \textbf{TN} & \textbf{FDR} & \textbf{TP} & \textbf{FP} & \textbf{TN} & \textbf{FDR} \\
\midrule
w/o FPF & 18 & 34 & 13 & 65.38\% & 26 & 27 & 18 & 50.94\% \\
\rowcolor{codegray}
\textbf{\framework} & 18 & 7 & 40 & 28.00\% & 26 & 5 & 40 & 16.13\%  \\
\bottomrule
\end{tabular}
\vspace{-0.15in}
\end{table}

On the other hand, to evaluate the contribution of false positive filter, since it is infeasible to determine the exact number of bugs in the client implementations to evaluate false negatives~\cite{ma2023loki}, we introduce false discovery rate (FDR) as $FDR=\frac{FP}{TP+FP}\times100\%$. In our evaluation, we define true positives (TP) as confirmed real bugs and false positives (FP) as false alarms. To count the number of TP, FP, and true negatives (TN), we manually analyzed all 136 responses (65 from EL client API, 71 from CL client API) and labeled them as TP, FP, or TN.
To reduce manual analysis effort, we generate one request for each the EL and  CL client APIs to assess the prevalence of false positives. 
As shown in Table~\ref{table:FDR:ablation}, the introduction of the FPF significantly reduces the FDR from 65.38\% to 28.00\% for Geth, and from 50.95\% to 16.13\% for Lighthouse. 

This improvement can be attributed to the FPF's capability to mitigate false positives arising from environmental discrepancies, filter out permissible variations in responses, and accurately identify semantically equivalent outputs across different clients. For instance, FPF recognizes error messages such as 'Unable to decode data' and 'Could not decode request body into...' as semantically equivalent, despite their syntactic differences.

\answerbox{\textbf{\noindent Answer for RQ4: } 
The ablation studies demonstrate that {\framework}'s key components significantly enhance its effectiveness, where fact-based semantics-aware generation (FSG) improves code coverage by up to 20.69\% by ensuring semantically valid test inputs, while the false positive filter (FPF) reduces the false discovery rate by 37.38\% for Geth and 34.82\% for Lighthouse, confirming that both components are essential for comprehensive and accurate API testing.
}

\section{Discussion}
In this section, we plan to conduct a thorough discussion from various perspectives.

\subsection{Implications}
We conduct a role-based implication discussion to demonstrate the broad impact and value of our work across different stakeholders in the Ethereum ecosystem.

\noindent{\textbf{Developers.}}
For Ethereum client developers, {\framework} provides tangible benefits by systematically identifying API implementation bugs that would otherwise remain undetected. Our comprehensive testing across all major clients has revealed 72 real-world bugs, with 90.27\% confirmed or fixed by developers, demonstrating the practical value of our approach. Each bug report includes detailed reproduction steps, API specifications, triggering requests, and comparative responses from other clients, enabling developers to quickly understand and resolve issues. 
The positive developer feedback, including acknowledgments, integration of our test cases, and expressed interest in our methodology, validates the utility of our approach.
We plan to integrate {\framework} into the Kurtosis Ethereum package~\cite{ether-package} as a plugin (\textit{e.g.,} dora~\cite{dora}) to streamline testing within existing development workflows.
This will allow development teams to continuously test their API implementations using Docker-based configurations, transforming {\framework} from a one-time testing tool into an ongoing quality assurance mechanism that can catch regressions and facilitates specification compliance throughout the development life-cycle.

\noindent\textbf{Ethereum Community.}
Client diversity is fundamental to Ethereum's decentralization and security model~\cite{diversity}, yet it introduces the risk of implementation inconsistencies that could threaten network stability. Our work directly strengthens this diversity by ensuring that different client implementations maintain functional equivalence. By testing 100\% of mainnet EL clients and 99.93\% of CL clients, {\framework} provides comprehensive coverage that validates the reliability of the entire client ecosystem.
The bugs we identified span all major clients, demonstrating that implementation inconsistencies are pervasive rather than isolated to specific teams or languages. By systematically detecting and reporting these issues, our work contributes to a more robust and reliable Ethereum infrastructure, enhancing the interoperability of EL and CL clients. This strengthens the foundation upon which Ethereum's client diversity strategy rests, making the network more resilient to single-point failures while maintaining consistency across implementations. 
Moreover, our findings highlight the real-world value of Ethereum infrastructure solutions like erpc~\cite{erpc}, Vero~\cite{vero}, and Vouch~\cite{vouch} in mitigating these implementation inconsistencies.

\noindent\textbf{Users.}
End users represent the ultimate beneficiaries of reliable Ethereum client APIs, as they depend on consistent and accurate responses for critical financial operations. As illustrated in our motivating example in Fig.~\ref{fig:mot_example}, even subtle API inconsistencies can lead to significant financial misunderstandings and damage trust in the entire ecosystem. Since most users interact with Ethereum through wallets like MetaMask~\cite{metamask} or third-party API providers like dRPC~\cite{drpc}, they have no visibility into which specific client implementation serves their requests and must rely on the underlying infrastructure to provide accurate data.
{\framework} directly benefits users by maintaining API consistency across client implementations. This consistency is particularly crucial for financial applications where accuracy is paramount, where users need confidence that their transaction history, account balances, and smart contract interactions are reported consistently regardless of the underlying client infrastructure. Our systematic testing helps provide the reliability that users expect from a mature financial platform, supporting the Ethereum Foundation's priority on improving user experience~\cite{ux}. 

\noindent\textbf{Researchers.} 
Our work opens new research directions in blockchain testing. Despite Ethereum's central role in the blockchain ecosystem, comprehensive API testing has been largely overlooked, with only limited tools exist. {\framework} represents the first systematic approach to testing both EL and CL client APIs, establishing a new standard for blockchain protocol testing.
The specification-guided testing methodology can be adapted to other blockchain platforms with well-documented APIs, potentially improving reliability across the broader blockchain ecosystem. Our approach of combining automated test generation with LLM-based false positive filtering provides a template for testing complex distributed systems. Furthermore, our comprehensive bug taxonomy and characterization provide valuable insights into common failure patterns in blockchain API implementations, informing future research in blockchain software engineering.

\subsection{Threats to Validity}
\label{subsec:threats}
Following established methodology~\cite{feldt2010validity}, we discuss threats to both internal and external validity that may affect the reliability and generality of our approach and evaluation results.

\noindent\textbf{Internal Validity.}
Internal validity concerns factors that may affect the accuracy of our experimental results and conclusions within the specific context of our study. We summarize these factors based on the workflow of {\framework}.

\textit{Specification Quality Dependency.} Our approach's effectiveness is tied to the quality of the underlying API specifications. The Ethereum client API specifications, while actively maintained~\cite{exec_api_github,beacon_api_github}, contain inherent limitations that affect our testing accuracy. First, certain API behaviors remain undefined, particularly corner cases and error handling scenarios. Developers are still actively discussing standardization of error codes~\cite{error_codes}. Second, the specifications themselves may contain bugs, as they are derived from higher-level specifications~\cite{exec-spec,consensus-specs} through human translation processes that can introduce flaws. Third, specifications provide syntactic requirements for parameters without semantic constraints or implementation details, creating ambiguities that lead to inconsistent implementations requiring developer discussions to resolve~\cite{spec_example}.

\textit{Semantic Test Generation.} Blockchain systems' inherent cryptographic dependencies present challenges for semantic test generation. Specifically, generating valid cryptographic parameters, such as hashes, signatures, and proofs, remains difficult. For instance, generating a valid signature requires the private key, the node state (to compute the appropriate signing domain), and compliance with the mathematical constraints of the BLS signature algorithm~\cite{BLS}.
This limitation compromises the semantic validity of certain test inputs, restricts the code coverage of \framework, and may cause bugs related to cryptographic parameter handling to be missed.

\textit{Choice of Black-box Testing.} While white-box techniques could offer deeper insights into internal program states, our decision to employ black-box testing is driven by the heterogeneity of the Ethereum ecosystem. As detailed in \S\ref{section:background:node}, there are over 11 distinct execution clients implemented in diverse programming languages (\textit{e.g.,} Go, Rust, Java, C\#) and maintained by independent teams. Adapting white-box instrumentation for every client would impose huge engineering overhead and restrict the scalability of our approach. Consequently, we prioritize the generality offered by black-box testing. By leveraging the standardized API, our method remains implementation-agnostic and can test any client adhering to the protocol, enabling a comprehensive assessment of the entire Ethereum ecosystem rather than a single client implementation.

\textit{Inherent Limitations of Differential Testing.} Despite its efficacy in uncovering logic and semantic discrepancies~\cite{mckeeman1998differential, jiangzenddiff, kim2023etherdiffer, petsios2017nezha, yang2021fluffy}, differential testing is fundamentally constrained by the oracle problem when all targets exhibit identical erroneous behavior in a specific scenario. To mitigate this, differential testing should be positioned as a complementary approach within a broader testing and verification suite~\cite{gulzar2019perception}. 
Specifically, metamorphic testing~\cite{segura2016survey,chen2018metamorphic} can also be employed when the oracle is unavailable by checking whether the software conforms to predefined metamorphic relations.
Unit and integration testing can be used to ensure base-level functional correctness. 
Furthermore, static analysis can be deployed to detect well-defined pattern-based vulnerabilities (\textit{e.g.,} memory leaks or data races). 
while runtime verification (\textit{e.g.,} assertions~\cite{rosenblum1995practical}) serves to catch internal invariant violations when outputs match.
Finally, although incurring higher overhead, formal verification methods~\cite{souyris2009formal,park2020end} provide rigorous guarantees and should be utilized for security-critical applications where differential testing may not suffice.

\textit{Limited Runtime Feedback.} Our test input generation relies solely on specifications without incorporating runtime feedback such as coverage data. This limitation restricts our ability to refine test inputs based on execution results. Additionally, coverage data collection presents challenges due to multi-language client implementations and requires node shutdown for data collection, causing significant testing downtime. These factors may limit the comprehensiveness of testing.

\textit{LLM-based Bug Identification.} A primary threat to internal validity stems from our reliance on LLMs to filter false positives during the bug identification process. 
We assume that LLMs can accurately interpret Ethereum client API specifications and correctly categorize response fields as \textit{must-identical}, \textit{may-divergent}, or \textit{must-divergent}.
However, due to the inherent stochastic nature of LLMs, there is a risk of misclassification, which may cause genuine bugs to be erroneously discarded as false positives. 
To mitigate this, we conducted manual verification of the labeled specifications to validate correctness. Nevertheless, we acknowledge the residual risk that the total bug count might be underestimated. 
To further counteract the unreliability and potential hallucinations of LLMs~\cite{10.1145/3703155}, a cross-validation method can be employed. 
For instance,  in \S\ref{subsec:bug}, multiple distinct LLM instances can be deployed to independently assess semantic equivalence. By strictly classifying a response pair as \texttt{semantically\_equivalent} only when all models reach a consensus, we can reduce the impact of individual model errors and enhance the robustness of the bug identification.

\noindent\textbf{External Validity.}
External validity addresses the generality of our findings beyond the specific Ethereum client APIs we tested.
Our approach focuses specifically on JSON-RPC and Beacon-API, which represent the most standardized and widely-supported API types in the Ethereum ecosystem. However, this may raise questions about the effectiveness of {\framework} when applied to other API types such as GraphQL~\cite{graphql} or custom client-specific APIs~\cite{customapi}. While our specification-guided approach should theoretically work with any well-documented API specifications, the effectiveness may vary depending on specification quality and the availability of semantic information for test input generation.

\subsection{Ethical Consideration}

For all identified inconsistencies, we responsibly reported issues to client developers and specification maintainers through Github issues.
For potentially security-sensitive issues, we followed the security policy specified by the respective client projects, reporting via their dedicated security channels.
Additionally, we submitted patches for the CL clients to help build a more robust and reliable ecosystem. Three fixes were accepted and merged by the developers.

\section{Related Work}

\noindent \textbf{Blockchain Client Testing.} 
Research on testing blockchain client API remains limited. EtherDiffer~\cite{kim2023etherdiffer} is the first differential testing tool for EL client API.
Hive~\cite{hive} is the official testing framework for running integration tests of Ethereum clients, which use \texttt{rpctestgen}~\cite{rpctg} for test input generation.
Moreover, prior studies~\cite{DBLP:conf/ndss/LiCLT0L21,infocom} have investigated security aspects of blockchain client API.
Li \textit{et al.} uncover Denial of Service (DoS) vulnerabilities in EL client API services~\cite{DBLP:conf/ndss/LiCLT0L21}.
Wang \textit{et al.}~\cite{infocom} propose a deanonymization attack targeting users of third-party Ethereum client API services.
Other studies have focused different components of blockchain clients, 
including the Ethereum Virtual Machine~\cite{opdiffer,fu2019evmfuzzer,maier2021NeoDiff}, consensus protocol implementations~\cite{fork_fuzz,ma2023loki,bft,chen2023tyr,yang2021fluffy}, mempool~\cite{MPFUZZ,deter,yaish}, state storage~\cite{nurgle,threadneck}, and concurrency bug detection~\cite{chord}. Specifically, EVM has also been studied through differential testing approaches.
EVMFuzzer \cite{fu2019evmfuzzer} is the first differential testing tool for EVM. NeoDiff \cite{maier2021NeoDiff} utilizes bytecode-level generation test input generation method to fuzz the EVM. OpDiffer~\cite{opdiffer} is an opcode-level differential testing framework for EVM. 
Consensus protocol testing has explored various techniques~\cite{fork_fuzz,ma2023loki,bft,chen2023tyr,yang2021fluffy}. Forky~\cite{fork_fuzz} applies differential testing to detect fork-handling bugs, while LOKI~\cite{ma2023loki} and Tyr~\cite{chen2023tyr} employ fuzzing strategies to identify memory-related and logical flaws in consensus mechanisms. ByzzFuzz~\cite{bft}, on the other hand, relies on random testing to uncover bugs in BFT implementations.
Research on Ethereum node performance and security has focused on examining the mempool and storage components. MPFUZZ~\cite{MPFUZZ}, DETER~\cite{deter}, and Yaish \textit{et al.}~\cite{yaish} specifically target DoS vulnerabilities in Ethereum mempools. Beyond mempools, NURGLE~\cite{nurgle} demonstrates a DoS attack targeting Ethereum's state storage. Additionally, ThreadNeck~\cite{threadneck} models the EL client as a set of threads to identify performance bottlenecks.

\noindent \textbf{API Testing.} 
In recent years, there has been a significant amount of research on API testing~\cite{api_survey}. Recent advancements have primarily focused on improving the effectiveness of automated testing tools, with a particular emphasis on RESTful API testing~\cite{evomaster,du2024vulnerability,atlidakis2019restler,lyu2023miner,kim2023enhancing,kim2024leveraging}. For example, EvoMaster is a well-known search-based fuzzing tool for testing RESTful APIs. Additionally, other notable efforts have explored testing for programming language-specific APIs~\cite{zhang2025rumono,takashima2021syrust,hu2020python}.

\noindent \textbf{Differential Testing.} Introduced by McKeeman in 1998~\cite{mckeeman1998differential}, differential testing has been widely applied in both academia and industry.
Beyond blockchain, differential testing has been applied to various domains, including detecting bugs in SSL/TLS implementations~\cite{SSL2014,SSL2023,petsios2017nezha}, compilers~\cite{guLLMBasedCodeGeneration2023,rust_compiler,feng2025finding}, runtime~\cite{lwdiff, drwasi, cao2024wasmaker, zhou2023wadiff, li2023pyrtfuzz,bernhard2022jit, chen2016coverage, chen2019deep}. 

\noindent \textbf{Semantic Testing.} 
Recent advancements have shifted the focus from syntactic structures to deeper semantic analysis~\cite{clark_semantic_2013,SemMT,pdiff,10136793}. 
To address the limitations of syntactic-based mutation, Clark \textit{et al.}~\cite{clark_semantic_2013} propose semantic mutation testing. Instead of generating mutants by altering source code, they mutate the semantic rules of the language. 
This allows for the detection of a different class of faults, particularly those caused by ambiguities within the C language specification.
PDiff~\cite{pdiff} adopts semantic summaries to capture the underlying logic of vulnerability patches for patch presence testing in open-source kernels.
SemMT~\cite{SemMT} leverages the semantic equivalence of logical relations and quantifiers to test the machine translation systems.
Ojdanic \textit{et al.}~\cite{10136793} underscores the necessity of evaluation of semantic similarity in fault seeding.

\noindent \textbf{Methodological Novelty.}
{\framework} distinguishes itself from existing works in three key aspects. 
First, unlike general-purpose API testing tools that rely on coarse-grained oracles (\textit{e.g.,} HTTP status codes)~\cite{api_survey}, {\framework} is the first to implement a specification-guided differential testing framework specifically for both Ethereum EL and CL client APIs, allowing it to capture deep semantic inconsistencies.
Second, distinct from prior semantic testing approaches that mutate inputs based on static rules, {\framework} leverages live blockchain states to improve the semantic validity of test inputs.
Third, classical differential testing suffers from substantial false positives due to the lack of determinism in distributed blockchain nodes. To address this limitation, we introduce a novel false positive filtering mechanism guided by specifications.
Consequently, {\framework} achieves higher code coverage and has detected a significant number of real-world bugs.
\section{Conclusion}
In this paper, we present {\framework}, a differential testing framework for Ethereum client API implementations.
We automatically generate syntactically valid and invalid test cases based on the API specifications and introduce fact-based semantics-aware request generation to further improve the semantical validity of test inputs.
To assist bug identification, we leverage heuristic-based methods and LLMs to filter false positives out.
Compared with state-of-the-art baselines, our evaluation illustrates that {\framework} can achieve up to 89.67\% code coverage improvement. 
Among 11 EL and CL clients, we detected 72 real-world API bugs, 91.38\% of which have been confirmed or fixed, including three bugs within the specifications.
{\framework} demonstrates significant real-world impact, with reported bugs receiving positive feedback such as acknowledgments, additional test cases, and expressed interest in the testing methodology, where one bug has even been brought up for discussion in the Ethereum official meeting.
\section*{Data-Availability Statement}
The artifact of {\framework} is available at Zenodo~\cite{code}.

\bibliographystyle{ACM-Reference-Format}
\bibliography{ref}

\end{document}